\documentclass[twocolumn,aps,superscriptaddress,nofootinbib,floatfix,linenumbers]{revtex4}

\usepackage{bm}
\usepackage{graphicx} 
\usepackage[utf8]{inputenc}
\usepackage[dvipsnames]{xcolor}
\usepackage{amsmath}
\usepackage{amssymb}

\newcommand{\tonde}[1]{\left({#1}\right)}
\newcommand{\quadre}[1]{\left[ {#1} \right]}

\begin{document}

\title{Far-from-equilibrium attractors with Full Relativistic Boltzmann approach in boost-invariant
and non-boost-invariant systems}

\author{Vincenzo Nugara}
\email{vincenzo.nugara@phd.unict.it}
\affiliation{Department of Physics and Astronomy, University of Catania, Via S. Sofia 64, I-95125 Catania}
\affiliation{INFN-Laboratori Nazionali del Sud, Via S. Sofia 62, I-95123 Catania, Italy}
\author{Salvatore Plumari}
\email{salvatore.plumari@dfa.unict.it}
\affiliation{Department of Physics and Astronomy, University of Catania, Via S. Sofia 64, I-95125 Catania}
\affiliation{INFN-Laboratori Nazionali del Sud, Via S. Sofia 62, I-95123 Catania, Italy}
\author{Lucia Oliva}
 
\affiliation{Department of Physics and Astronomy, University of Catania, Via S. Sofia 64, I-95125 Catania}
\affiliation{INFN Sezione di Catania, Via S. Sofia 64, I-95123 Catania, Italy}

\author{Vincenzo Greco}
\affiliation{Department of Physics and Astronomy, University of Catania, Via S. Sofia 64, I-95125 Catania}
\affiliation{INFN-Laboratori Nazionali del Sud, Via S. Sofia 62, I-95123 Catania, Italy}
\email{vincenzo.greco@dfa.unict.it}

\begin{abstract}
We study the universal behavior associated with a Relativistic Boltzmann Transport (RBT) approach with the full collision integral in 0+1D conformal systems.
We show that all momentum moments of the distribution function exhibit universal behavior. Furthermore, the RBT approach allows to calculate the full distribution function, showing that an attractor behavior is present in both the longitudinal and transverse momentum dependence.
We compare our results to the far-from-equilibrium attractors determined with other approaches, such as kinetic theory in Relaxation Time Approximation (RTA) and relativistic hydrodynamic theories, both in their viscous (DNMR) an anisotropic (aHydro) formulations, finding a very similar evolution, but an even faster thermalization in RBT for higher order moments.

For the first time, we extended this analysis also to study the attractor behavior under a temperature-dependent viscosity $\eta/s(T)$, accounting also for the rapid increase toward the hadronic phase. 
We find that a partial breaking of the scaling behavior with respect to $\tau/\tau_{eq}$ emerges only at $T \approx T_c$ generating a transient deviation from attractors; interestingly this in realistic finite systems may occur around the freeze-out dynamics.
Finally, we investigate for the first time results beyond the boost-invariant picture, finding that also in such a case the system evolves toward the universal attractor. In particular, we present the forward and pull-back attractors at different space-time rapidities including rapidity regions where initially the distribution function is even vanishing.

\end{abstract}

\maketitle

\section{Introduction}

Relativistic heavy-ion collisions provide a unique laboratory for probing the fundamental properties of nuclear matter under extreme conditions, studying the formation and evolution of the Quark-Gluon Plasma (QGP). The dynamics of this deconfined state of strongly interacting matter created in ultra-Relativistic Heavy Ion Collisions (uRHICs) can be successfully described by means of both macroscopic hydrodynamics theories \cite{Romatschke:2017ejr, Florkowski:2017olj, Heinz:2013th, Denicol:2012cn} and microscopic kinetic transport approaches \cite{Ruggieri:2013bda, Ruggieri:2013ova,  Ruggieri:2015yea, Plumari:2015sia, Plumari:2019gwq, Xu:2004mz, Uphoff:2014cba, Cassing:2009vt, Bratkovskaya:2011wp, Soloveva:2021quj, Kurkela:2018vqr, Oliva:2022rsv}. 
Hydrodynamics, especially in its viscous and anisotropic formulations, is able to describe the evolution of macroscopic quantities in different scenarios and also in situations where the dynamics is far from the equilibrium, leading to an agreement with experimental measurements on hadron transverse momentum spectra and anisotropic flows in uRHICs.
This is remarkable considering that the early stage of the collision is highly anisotropic and out of equilibrium and therefore in a regime beyond the expected applicability of hydrodynamics. 
Besides uRHICs, in recent years small systems, such as proton-proton ($pp$) and proton-nucleus collisions ($pA$), have attracted great attention, since a collective behavior like that observed in heavy-ion collisions has been found also in high-multiplicity events of small colliding systems, rising the question of the possible formation of QGP \cite{Yan:2013laa, Romatschke:2015gxa, Shen:2016zpp, Mantysaari:2017cni}.
Even more unexpectedly, the hydrodynamic model has been successful in describing collective observables in high-multiplicity events of small colliding systems, which most likely remain out of equilibrium throughout their dynamical evolution \cite{Zhao:2022ugy, Oliva:2022rsv}.
Therefore, the comprehension of the thermalization mechanisms and timescales of systems which are far from equilibrium is of
fundamental importance in the context of uRHICs and even more in the case of small collision 
systems.
This has led to recent advances in the theoretical framework, bringing to light a striking phenomenon: the emergence of universal behavior in the context of relativistic hydrodynamics, kinetic theory, classical Yang-Mills equations as well as in AdS/CFT calculations (hydrodynamics \cite{Heller:2015dha, Strickland:2017kux,  Chattopadhyay:2019jqj, Jaiswal:2019cju,Blaizot:2019scw, Alalawi:2020zbx, Heller:2020anv}, Effective Kinetic Theory (EKT) \cite{Kurkela:2018vqr, Kurkela:2019set, Almaalol:2020rnu}, Relaxation Time Approximation (RTA) kinetic theory \cite{Behtash:2017wqg, Blaizot:2017ucy,  Strickland:2018ayk, Heller:2018qvh, Kamata:2020mka}, small-angle kinetic theory \cite{Tanji:2017suk, Brewer:2022vkq}, BAMPS \cite{Ambrus:2021sjg}, Yang-Mills equations \cite{Berges:2013eia,Berges:2013fga}, AdS-CFT \cite{Kurkela:2019set, Heller:2011ju}).
Indeed, the far-from-equilibrium evolution of macroscopic quantities, irrespective of the initial conditions, exhibit universal scaling, denoted as non-equilibrium attractors, endowed with the noteworthy ability to capture the long-term collective behavior of the fluid system.

In this paper we extend this analysis in the context of kinetic theory going beyond the RTA and solving the full collision integral of the Relativistic Boltzmann Transport (RBT) equation.
The approach of the system toward the equilibrium has been investigated, especially within calculations based on RTA kinetic theory and anisotropic hydrodynamics (aHydro) \cite{Martinez:2010sc, Florkowski:2010cf, Almaalol:2018jmz}, by looking at the evolution of the momentum moments of the phase-space distribution function. 
We have computed such moments by means of simulations with the full RBT approach with different values of the shear viscosity over entropy density ratio $\eta/s$ as well as different initial conditions. We have studied the attractor of the RBT and we have compared our findings with those of RTA, aHydro and viscous 
Hydrodynamics (vHydro).
All previous studies have shown the existence of attractors in the case of longitudinal boost-invariant expansion for a system evolving with fixed $\eta/s$.
Apart from this there are two novelties in our study. On one hand we consider a more realistic case, investigating the impact of the temperature-dependent $\eta/s$ on the attractor behavior; on the other hand, we discuss for the first time the universal behavior of the system at large rapidity, studying also the case where the longitudinal boost invariance is dynamically broken. In the former case the universal scaling is still present at
large scaled times, despite a temporary breaking, which emerges at intermediate $\tau$ corresponding to lower $T(\tau)$ and
higher specific viscosity values. In the latter case, we find a universal behavior even if the system explicitly breaks boost-invariance, by imposing an initial finite distribution in $\eta$.

The paper is structured as follows.
In Sec. \ref{sec:RBT} we explain in detail our full Boltzmann approach, while the main ingredients of the models to which we compare (RTA, vHydro and aHydro) are presented in Sec. \ref{sec:models}.
In Sec. \ref{sec:iso_therm} we show the time evolution of the distribution function moments in the RBT approach at fixed viscosity in comparison with the free-streaming case and with the results from the other theoretical frameworks. Moreover, we discuss the role of the inverse collision rate as the timescale that determines the emergence of the attractor behavior in RBT.
In Sec. \ref{sec:pullback_forward} we delve into the search for universal scaling in RBT, by showing pull-back and forward attractors for the momentum moments as well as for the time evolution of the full distribution function.
In Sec. \ref{sec:Tdep_visco} we extend the analysis of the momentum moments for different parametrizations of $\eta/s(T)$.
Finally, in Sec. \ref{sec:no_boostinv} we analyze the attractor behavior at different rapidities and in the case of dynamical breaking of longitudinal boost-invariance.
The conclusions of our work are summarized in Sec. \ref{sec:concl}.\newline

\section{Full Boltzmann Approach}\label{sec:RBT}
In this work we employ the Relativistic Boltzmann Transport (RBT) approach to study the evolution of the fireball created in uRHICs.   
In our simulation we employ a relativistic transport code developed in these years to perform studies of the quark-gluon plasma dynamics in uRHICs for different systems from RHIC to LHC energies 
\cite{Ferini:2008he, Plumari:2012ep, Ruggieri:2013bda, Scardina:2012mik, Ruggieri:2013ova, Puglisi:2014sha, Scardina:2014gxa, Plumari:2015sia, Plumari:2015cfa, Scardina:2017ipo, Plumari:2019gwq, Sun:2019gxg, Sambataro:2020pge, Gabbana:2019uqv}.
In our approach we describe QGP matter with an on-shell one-particle
distribution function $f$, depending on space-time coordinates $x^{\mu}=\left(t,\mathbf{x}\right)$
and 4-momentum $p^{\mu}=(p^0,\mathbf{p})$, being $p^{0}=E_{\mathbf{p}}\equiv\sqrt{m^{2}+\mathbf{p}^{2}}$. The space-time evolution of $f$ is
governed by the following RBT equation
\begin{equation}
p^{\mu}\partial_{\mu}f(x,p)=C\left[f(x,p)\right]_{\mathbf{p}},
\label{eq:RBT}
\end{equation}
being $\partial_{\mu}\equiv\partial/\partial x^{\mu}$
the gradient with respect to space-time coordinates $x^{\mu}$.
In this work we are interested in the conformal limit, hence the particle mass is zero.
The quantity $C\left[f\right]_{\mathbf{p}}$ is the collision integral which, considering only binary collisions as we do in this paper, is given by
\begin{gather}
C\left[f\right]_{\mathbf{p}} =
\intop\frac{\text{d}^{3}p_{2}}{2E_{\mathbf{p}_{2}}\left(2\pi\right)^{3}}
\intop\frac{\text{d}^{3}p_{1'}}{2E_{\mathbf{p}_{1'}}\left(2\pi\right)^{3}}
\int\frac{\text{d}^{3}p_{2'}}{2E_{\mathbf{p}_{2'}}\left(2\pi\right)^{3}}\nonumber \\
\times\left(f_{1'}f_{2'}-f_{1}f_{2}\right)  \left|\mathcal{M}\right|^{2}\delta^{\left(4\right)}\left(p_{1}+p_{2}-p_{1'}-p_{2'}\right),
\end{gather}
where $f_i=f(p_i)$, $\mathcal{M}$ denotes the transition amplitude for the elastic processes which is directly linked to the differential cross section $|{\cal M}|^2=16 \pi\,s\,(s-4M^2) d\sigma/dt$ with $s$ the Mandelstam invariant.
The symbol $\left[\cdot\right]_{\mathbf{p}}$ indicates that $C$
is a functional of $f$ with respect to 3-momentum. 

Numerically, Eq. \eqref{eq:RBT} is solved by discretizing the space-time in cells and using the so called test particle method \citep{Wong:1982zzb}. Within this method, the one-particle distribution function is sampled by considering a finite number $N$ of point-like test particles so that the phase-space distribution can be written as

\begin{equation}
f\left(t,\mathbf{x},\mathbf{p}\right)=\frac{1}{N_{\text{test}}}\sum_{i=1}^{N}\delta^{\left(3\right)}\left(\mathbf{x}-\mathbf{x}_{i}\left(t\right)\right)\delta^{\left(3\right)}\left(\mathbf{p}-\mathbf{p}_{i}\left(t\right)\right),\label{eq:f_test}
\end{equation}
where $N_{test}$ is the number of test particles per real particle and $\mathbf{x}_{i}\left(t\right)$ and $\mathbf{p}_{i}\left(t\right)$ are the position and 3-momentum of the $i-$th test particle at time $t$
respectively. 
It can be shown, by using Liouville theorem \cite{LANG1993391}, that Eq. \eqref{eq:f_test} is a solution of the RBT equation provided 
that every test particle is propagated according to the  relativistic Hamilton equations of motion
\begin{align}
 & \dot{\mathbf{x}}_{i}\left(t\right)=\frac{\mathbf{p}_{i}\left(t\right)}{p_{i}^0\left(t\right)}\label{eq:Particle position equation of motion}\\
 & \dot{\mathbf{p}}_{i}\left(t\right)=coll.
 \label{eq:eq_motion}
\end{align}
These equations are solved with a 4th-order Runge-Kutta integration method on a space-time grid.
The term $coll.$ on the RHS of the Hamilton equation indicates the effect of the collision integral in modifying particle momenta. 
In our calculation this term is included using the so-called stochastic algorithm for the collision integral and numerically solved as in Ref. \cite{Xu:2004mz,Ferini:2008he}.\\

{The code is written in C language and has been optimized for High Performance Computing (HPC). Space is discretized in cells with fixed size in $x$, $y$ and $\eta$, thus their length in $z$ will be expanding linearly in $t$ according to the relation:
$$ \Delta z_j = z_{j+1} - z_{j} = t (\tanh \eta_{j+1} - \tanh \eta_j), $$
where $j$ is the cell index.
This allows us to have an approximately constant number of test particles per cell during the whole evolution. As far as the simulations whose results will be showed in the following, we fixed $\Delta x = 0.4$ fm, $\Delta y = 0.4 $ fm and $\Delta \eta = 0.04$. \\ In order to preserve causality, the time discretization is strongly bounded by the space one: since particles within the same cell are allowed to collide and exchange momentum, the maximum length within the cell must be smaller than $\Delta t$.
For our simulations, we employ about 1500 $-$ 3000 time-steps, depending on the final time $t_{fin}$. Finally, we are able to simulate systems with up to $3\cdot 10^8$ total particles, achieving quite good statistics. Results are extracted by considering a selection in $\eta$, whose thickness is equal to $\Delta\eta = 0.04$, unless otherwise specified. The setup described above has been verified to guarantee convergence and stability of the Relativistic Boltzmann Equation.\\
As far as the running time, we checked it being linear with the number of total particles and time steps: for values of 2000 time steps and $10^8$ total particles as used in this work, a simulation requires 100 core-hours. More realistic simulations to which we aim, including event-by-event fluctuations that require at least $10^4$ events, will require about 1M core-hours.}

In standard kinetic theory one fixes microscopic cross sections and, consequently, the specific viscosity $\eta/s$ emerges as an intrinsic macroscopic quantity. 
In contrast, in our approach we connect the scattering cross section $\sigma$ to a given $\eta/s$, in order to have a more direct comparison to hydrodynamic calculation at fixed $\eta/s$.
An analytical relation between $\eta$, $T$ and $\sigma$ is obtained within the Chapman-Enskog approximation:
\begin{eqnarray}
\eta=1.2\frac{T}{\sigma} \label{eq:eta_CE}
\end{eqnarray}
As shown in Ref.~\cite{Plumari:2012ep} the above expression for the shear viscosity $\eta$ is in quite good agreement with the Green-Kubo formula.
Since we are interested in fixing $\eta/s$, we assume local thermodynamic equilibrium and use the relation between the entropy density $s$ and the particle density $n$ in the Local Rest Frame (LRF), i.e., $s=n \big( 4 -\ln{\Gamma}\big)$, where $\Gamma$ is the fugacity, and rewrite equation (\ref{eq:eta_CE}) as
\begin{equation}
\sigma=1.2\frac{T}{n\left(4-\ln{\Gamma}\right)\eta/s}.
\label{eq:sigma}
\end{equation}

The validity of this probabilistic nature of the collision integral strongly depends on the value of $\eta/s$. In fact, for $\eta/s\rightarrow 0$ the cross section becomes non-physically large, leading to numerical
instabilities. However, it has been shown in \citep{Gabbana:2019uqv} that
this approach is able to describe the dynamical evolution also at very low values, even below $\eta/s=0.05$ which is smaller than the 
conjectured minimal bound of $1/4\pi$ for the QGP \cite{Kovtun:2004de}.\\
In order to extract the local thermodynamic quantities, one has to compute the energy density $\varepsilon$ and particle density $n$ in the LRF. In the definition of the LRF there is some arbitrariness; following what is usually done in literature, we define the LRF as the Landau frame \cite{Denicol:book}. The energy density $\varepsilon$ and the fluid four-velocity $u_{\mu}$ are defined as the largest eigenvalue and the associated eigenvector of the energy-momentum tensor:
\begin{equation}\label{eq:Landau_frame}
    T^{\mu\nu} u_{\nu} = \varepsilon u^{\mu},
\end{equation}
where $u^\mu =\gamma\ (1,\boldsymbol{\beta})$, with $\boldsymbol{\beta}$ being the 3-velocity of the fluid element and $\gamma = 1/\sqrt{1-\boldsymbol{\beta}^2}$ the corresponding Lorentz factor. In kinetic theory $T^{\mu\nu}$ is given by
\begin{equation}
    T^{\mu\nu}(x)= \int dP \, p^{\mu} p^{\nu} f(x,p),
    \label{eq:Tmunu_kinetic}
\end{equation}
with $dP=d^3p/\left((2\pi)^3p^0\right)$ being the integration measure in the momentum space.
The local particle density is given by the expression
\begin{equation} \label{eq:densityLRF}
    n = n^\mu u_\mu,
\end{equation}
where the four-current $n^\mu$ in kinetic theory reads
\begin{equation}\label{eq:Nmu_kinetic}
    n^\mu = \int dP \, p^\mu f(x,p).
\end{equation}
In our approach, we only consider elastic $2\leftrightarrow 2$ collisions, therefore the number of particles is conserved, bringing the system away from chemical equilibrium, since there is no
process able to balance the chemical potential gradients
originated by the fluid expansion. This leads to a fugacity $\Gamma(\tau)\ne 1$. In order to describe the thermalization of the system, we must take into account the fugacity in imposing two Landau matching conditions for the energy and particle densities.
These conditions define the effective time-dependent temperature $T$ and fugacity $\Gamma$ as 
\begin{equation} \label{eq:matching_conditions}
    T(\tau) \equiv\frac{\varepsilon (\tau)}{3\,n(\tau)}, \qquad
    \Gamma(\tau)\equiv\frac{n(\tau)}{d \,T(\tau)^3/\pi^2},
\end{equation}
where $d$ is the number of degrees of freedom of the system, assumed to be $d=1$ henceforth.

We compare the results obtained with the RBT approach with other models, which will be briefly described below: the solution of the Boltzmann equation in Relaxation Time Approximation (RTA), {second-order dissipative viscous hydrodynamics} and anisotropic hydrodynamics. In order to perform such a comparison we analyze the moments of the distribution function in the various models. Following Ref. \cite{Strickland:2018ayk}, we define the momentum moments of the distribution function for on-shell particles as
\begin{equation}\label{eq:momentum_moments}
\mathcal{M}^{mn}(\tau)=\int dP \, (p\cdot u)^n \, (p\cdot z) ^{2m} \, f(\tau,p).
\end{equation}
It is useful to define the normalized moments $\overline M^{mn}=\mathcal{M}^{mn}/\mathcal{M}^{mn}_{eq}$ \cite{Strickland:2019hff}, where the moments are rescaled by their corresponding equilibrium values $\mathcal{M}^{nm}_{eq}$.
In the case of massless Boltzmann distribution with particle number conservation the equilibrium moments are given by {\cite{Strickland:2019hff}:
\begin{equation}\label{eq:MeqBoltz}
    \mathcal{M}^{nm}_{eq} (\tau) = \frac{(n+2m+1)!\,\Gamma(\tau)\, T^{n+2m+2}(\tau)}{2\pi^2 (2m+1)},
\end{equation}
which are exactly the moments in Eq. \eqref{eq:momentum_moments} computed when $f_{eq} (\tau,p) = \Gamma(\tau) \exp (-p\cdot u(\tau)/T(\tau))$, with $T$ and  $\Gamma$ being the effective time-dependent temperature and fugacity, defined in Eq. \eqref{eq:matching_conditions}.

\section{Models}\label{sec:models}

In this section we present the different frameworks used in the paper: Relaxation Time approximation (RTA), viscous hydrodynamics and anisotropic Hydrodynamics (aHydro). In order to make the paper self-consistent we only present the main points needed for the discussion and we refer the reader to the respective references for more details \cite{Strickland:2019hff, Denicol:2010xn, Almaalol:2018jmz}. 
In the following subsections, the initial condition for the distribution function is given by the Romatschke-Strickland distribution \cite{Romatschke:2003ms}, that is a spheroidally deformed thermal initial condition for an ideal gas:
\begin{equation}\label{eq:frs}
    f_0 (p) = \gamma_0 \exp \tonde{ - \frac{\sqrt{ p_T^2 + (1+\xi_0)(p\cdot z)^2 }}{\Lambda_0} },
\end{equation}
where the parameters $\Lambda_0$ and $\gamma_0$ are computed to fix the initial energy and particle density, while $\xi_0$ quantifies the system momentum anisotropy. It is also useful for the discussion to introduce the elliptical anisotropy parameter $\alpha_0 = (1+\xi_0)^{-1/2}$.
Notice that, as will be shown in Sec. \ref{subsec:aHydro} the isotropic Boltzmann distribution is recovered when $\xi_0=0$; in this case the parameter $\Lambda_0$ corresponds to the usual initial temperature $T_0$ and $\gamma_0$ to the standard fugacity $\Gamma_0$.

\subsection{Relaxation time approximation with number-conserving kernel}

The RBT Eq. \eqref{eq:RBT} can be solved exactly in the case of Bjorken flow in the relaxation time approximation (RTA), namely assuming that the distribution function converges to the equilibrium distribution function $f_{eq}$ within a time-scale $\tau_{eq}$. Therefore, the full collision integral is replaced by a simplified expression and Eq. \eqref{eq:RBT} becomes:
\begin{equation}\label{eq:RTA}
    p^\mu \partial_\mu f(x,p) = - \frac{p\cdot u}{\tau_{eq}} (f-f_{eq}),
\end{equation}
where $u$ is the fluid four-velocity and $f_{eq} (\tau,p) = \Gamma(\tau) \exp (-p\cdot u(\tau)/T(\tau))$, with $T$ and  $\Gamma$ being the effective time-dependent temperature and fugacity, defined as in Eq. \eqref{eq:matching_conditions}.

The local energy density $\varepsilon$ and particle density $n$ are obtained through Eqs.\eqref{eq:Landau_frame} and \eqref{eq:densityLRF}, where $T^{\mu\nu}$ and $n^{\mu}$ are computed with Eqs.\eqref{eq:Tmunu_kinetic} and \eqref{eq:Nmu_kinetic} respectively.

Proceeding as in Ref. \cite{Strickland:2019hff}, we solve the two following coupled integro-differential equations by the iterative method in a discrete lattice in $\tau$:
\begin{subequations}
\begin{gather}
    \begin{split}
        \Gamma(\tau) T^4(\tau) &= D(\tau,\tau_0) \Gamma_0 T_0^4 \frac{\mathcal H^{20}(\alpha_0 \tau_0 /\tau)}{\mathcal H^{20}(\alpha_0)} \\ & + \int_{\tau_0}^\tau \frac{d\tau'}{2\tau_{eq}(\tau')} D(\tau,\tau') \Gamma(\tau') T^4(\tau') \mathcal H^{20}\tonde{\frac{\tau'}{\tau}},\\
    \end{split}\\
    \begin{split}
         \Gamma(\tau)T^3(\tau')& = \frac{1}{\tau} \Big[ D(\tau,\tau_0) \Gamma_0 T_0^3 \tau_0 \\ & + \int_{\tau_0}^\tau \frac{d\tau'}{\tau_{eq}(\tau')} D(\tau,\tau') \Gamma(\tau') T^3(\tau') \tau'\Big],
    \end{split}
\end{gather}
\end{subequations}
where the relaxation time $\tau_{eq}(\tau)$ is given by the following ansatz \cite{Denicol:2010xn}:
\begin{equation}\label{eq:tau_RTA}
    \tau_{eq}(\tau) = \frac{5\eta/s}{T(\tau)},
\end{equation}
while the auxiliary functions are
\begin{subequations}
\begin{gather}
        D(\tau_2,\tau_1) = \exp \quadre{ \int_{\tau_1}^{\tau_2} \frac{d\tau''}{\tau_{eq}(\tau'')} },\\
    \mathcal H^{nm} (y) = \frac{2 y^{2m+1}}{2m+1} {}_2F_1\tonde{ \frac 12 +m , \frac{1-n}{2}; \frac 32 + m; 1-y^2 }, \label{eq:H_nm_strick}  
\end{gather}
\end{subequations}
with ${}_2F_1(a,b; c; z)$ being the ordinary hypergeometric function.\\

Finally, the moments \eqref{eq:momentum_moments} of the distribution function can be computed given $T(\tau)$ and $\Gamma(\tau)$ as
\begin{multline}
    M^{nm}(\tau) = \frac{(n+2m+1)!}{(2\pi)^2}\\ \times
    \bigg [
    D(\tau,\tau_0) \alpha_0^{n+2m-2} T_0^{n+2m+2}\Gamma_0 \frac{ \mathcal H^{nm} (\alpha\tau_0/\tau) }{ [\mathcal H^{20}(\alpha_0)/2]^{n+2m-1} } \\
    + \int_{\tau_0}^\tau \frac{d\tau'}{\tau_{eq}(\tau')} D(\tau',\tau') \Gamma(\tau') T^{n+2m+2}(\tau') \mathcal H^{nm} \tonde{ \frac{\tau'}{\tau} } \bigg ].
\end{multline}

\subsection{Viscous Hydrodynamics}

Differently from the transport kinetic approaches where the dynamics is followed by means of evolution equations for the microscopic particle distribution function, as seen in the previous sections, in relativistic hydrodynamics the space-time evolution of the medium is determined by conservation equations of macroscopic quantities, i.e. the particle 4-current $n^\mu$ and the energy-momentum tensor $T^{\mu\nu}$:
\begin{gather}
    \partial_\mu n^\mu =0,
    \label{eq:evol_nmu}\\
    \partial_\mu T^{\mu\nu} =0.
    \label{eq:evol_Tmunu}
\end{gather}
The energy-momentum tensor for a viscous fluid is related to the energy density $\varepsilon$, the equilibrium pressure $P$ in the LRF and the viscous corrections $\Pi$ and $\pi^{\mu\nu}$ by the formula
\begin{equation}
    T^{\mu\nu}=\varepsilon u^{\mu}u^{\nu}- \Delta^{\mu\nu} (P + \Pi) + \pi^{\mu\nu},
  \label{eq:Tmunu_hydro}
\end{equation}
with $\Delta^{\mu\nu}=g^{\mu\nu}-u^\mu u^\nu$ being the projector on the subspace orthogonal to $u^\mu$ and $g^{\mu\nu}=\mathrm{diag}(1,-1,-1,-1)$ being the metric tensor.
We consider the case of a conformal system ($m=0$), hence the bulk viscous pressure is $\Pi=0$. In the case of Bjorken flow, the conservation equation Eq. \eqref{eq:evol_Tmunu} can be written as a system of two coupled evolution equations for $\varepsilon$ and $\pi$ \cite{Denicol:2010xn, Denicol:2012cn}:
\begin{subequations} \label{eq:vHydro}
\begin{gather}
\label{eq:vHydro_eps}
    \partial_\tau \varepsilon = -\frac{1}{\tau} (\varepsilon + P -\pi),\\
    \partial_\tau \pi = - \frac{\pi}{\tau_\pi} + \frac 43 \frac{\eta}{\tau_\pi \tau} - \lambda \frac{\pi}{\tau}, \label{eq:vHydro_pi}
    \end{gather}
\end{subequations}
where the shear viscous pressure is $\pi=\sqrt{2\pi^{\mu\nu}\pi_{\mu\nu}/3}$ and the shear relaxation time is $\tau_\pi=\tau_{eq}=5(\eta/s)/T$.
The transport coefficient $\lambda$ according to DNMR theory in 14-moment approximation is computed as 124/63. The initial anisotropy $\xi_0$ is chosen by fixing the initial ratio $\pi/\varepsilon$ according to the formula \cite{Strickland:2018ayk}:
\begin{equation}
    \frac{\pi}{\varepsilon}(\xi_0)= \frac 13 \tonde{1 - \frac{\mathcal R(\xi_0)}{\mathcal R_L(\xi_0)}},
\end{equation}
where
\begin{gather}
    \label{eq:R(xi)}
     \mathcal R(\xi)=\frac 12 \quadre{ \frac{1}{1+\xi} + \frac{\arctan \sqrt{\xi}}{\sqrt{\xi}} },\\
     \mathcal R_L(\xi) = \frac{3}{\xi} \quadre{ \frac{(\xi+1) \mathcal R(\xi) -1 }{\xi +1} }.
\end{gather}
The results showed in this paper are obtained by numerically solving Eqs. \eqref{eq:vHydro} making use of a 4th-order Runge-Kutta method.
In the microscopic derivation of the viscous hydrodynamic equations \eqref{eq:vHydro} \cite{Denicol:book}, an ansatz for the deviation $\delta f$ of the distribution function with respect to the equilibrium value $f_{eq}$ must be given: $f = f_{eq} + \delta f$. In the DNMR 14-moments approximation \cite{Denicol:2010xn}:
\begin{equation}
    \delta f_{\text{DNMR}} = \frac{3}{16} \frac{f_{eq} (T)}{T^2}[(p\cdot u)^2 - 3 (p\cdot z)^2] \frac{\pi}{\varepsilon}.
    \label{eq:df_IS}
\end{equation}
Notice that in the Chapman-Enskog approach \cite{Romatschke:2011qp} $\delta f_{\text{CE}}/f_{eq} \sim p/T$, while $\delta f_{\text{DNMR}}/f_{eq} \sim (p/T)^2$. These trends can be compared with the outcomes of RBT transport theory investigated by some of the authors a few years ago \cite{Plumari:2015sia}, which show that, in the case of isotropic cross section, $\delta f/f_{eq} \sim p_T ^ {0.98}$, hence essentially similar to Chapman-Enskog ansatz.\\
Inserting $f = f_{eq} + \delta f$ with the expression for $\delta f$ given in Eq. \eqref{eq:df_IS}, the momentum moments of Eq. \eqref{eq:momentum_moments} normalized to their equilibrium value become: 
\begin{gather}\label{eq:vhydro_moments}
    \overline M^{nm}_{\text{DNMR}} = 1- \frac{3m(n+2m+2)(n+2m+3)}{4(2m+3)} \frac{\pi}{\varepsilon}.
\end{gather}

\subsection{Anisotropic hydrodynamics with number-conserving kernel}
\label{subsec:aHydro}

Starting from a set of momentum moments of the RTA Boltzmann equation (\ref{eq:RTA}) and assuming the Romatschke-Strickland form for the distribution function
\begin{equation}
    f(p, \tau) = \gamma (\tau) \exp \tonde{ - \frac{\sqrt{ p_T^2 + (1+\xi(\tau))(p\cdot z)^2 }}{\Lambda (\tau)} },
\end{equation}   
it is possible to derive  a system of three coupled ODEs \cite{Alqahtani:2017mhy} which describe the evolution of the system in terms of three macroscopic functions: $\gamma$, that is related to the effective fugacity, $\Lambda$, that in linked to the effective temperature, and $\xi$, that is connected to the pressure anisotropy $P_L/P_T$.

The zeroth- and the first-moment equations account for the number and energy-momentum conservation Eqs. \eqref{eq:evol_nmu}-\eqref{eq:evol_Tmunu}.
The second-moment equation is the difference between the $zz$ projection and one third of the sum of the $xx$, $yy$ and $zz$ projections of the equation:
\begin{equation}\label{eq:2nd_moment}
    \partial_\lambda I^{\lambda \mu\nu} = \frac{1}{\tau_{eq}} (u_\lambda I_{eq}^{\lambda \mu \nu} - u_\lambda I^{\lambda\mu\nu})
\end{equation}
where
\begin{equation*}
    I^{\lambda\mu\nu} = \int dP \, p^\lambda p^\mu p^\nu f.
\end{equation*}
By writing Eqs.\eqref{eq:evol_nmu}, \eqref{eq:evol_Tmunu} and \eqref{eq:2nd_moment} in terms of $\Lambda$, $\gamma$ and $\xi$ one gets the three coupled ODEs:   
\begin{equation}
    \label{eq:aHydro}
    \begin{split}
    & \partial_\tau \log \gamma + 3 \partial_\tau \log \Lambda - \frac 12 \frac{\partial_\tau \xi}{1+\xi} + \frac{1}{\tau} =0,\\
    & \partial_\tau \log \gamma + 4 \partial_\tau \log \Lambda + \frac{\mathcal R'(\xi)}{\mathcal R(\xi)} \partial_\tau \xi = \frac{1}{\tau} \quadre{ \frac{1}{\xi(1+\xi) \mathcal(\xi)} - \frac{1}{\xi} -1 },\\
    & \partial_\tau \xi - \frac{2(1+\xi)}{\tau} + \frac{\xi (1+\xi)^2 \mathcal R^2(\xi)}{\tau_{eq}}=0,
    \end{split}
\end{equation}
where the relaxation time is given by Eq. \eqref{eq:tau_RTA} and the effective temperature $T$ and fugacity $\Gamma$ are related to $\Lambda$ and $\gamma$ via the function $\mathcal R(\xi)$ defined in Eq. \eqref{eq:R(xi)}:
\begin{gather*}
    T= \mathcal R(\xi) \sqrt{1 + \xi} \Lambda,\\
    \Gamma = \frac{\gamma}{(1+\xi)^2 \mathcal R^3(\xi)}.
\end{gather*}
Notice that, if $\xi=0$, $\Lambda$ and $\gamma$ reduce respectively to $T$ and $\Gamma$.

The expressions for the normalized moments are:
\begin{equation}\label{eq:ahydro_moments}
    \overline M^{nm}_{\text{aHydro}} (\tau) = (2m+1) (2\alpha)^{n+2m-2} \frac{\mathcal H^{nm} (\alpha) }{ [\mathcal H^{20} (\alpha)]^{n+2m-1} },
\end{equation}
where $\mathcal H^{nm}$ are the functions defined in Eq. \eqref{eq:H_nm_strick}.

\section{Isotropization and thermalization}\label{sec:iso_therm}

In this work we are interested in studying the one-dimensional evolution along the longitudinal direction; therefore, we consider a box expanding in the longitudinal direction itself. In the transverse plane we consider a square of $5.2\times5.2$ fm$^2$, with periodic boundary conditions: when the transverse coordinate of a test particle exceeds the limit of the square, it enters in the opposite direction of the same amount it exceeded; particles keep their own momentum when re-entering into the square. This corresponds to simulate an infinite system in the transverse direction, which is equivalent to a purely 1D expanding system. 

In our calculations, we fix $\Lambda_0$ in order to have the initial energy density corresponding to a massless gas at a given temperature $T_0$. The results presented in the following, unless otherwise specified, are obtained with $T_0=0.5$ GeV.
Our initial conditions for the phase-space distribution function are given by the Romatschke-Strickland distribution in Eq. \eqref{eq:frs}.
The parameter $\gamma_0$ is fixed in order to have a system with initial fugacity $\Gamma_0=1$.
We will show results for different values of the anisotropy parameter $\xi_0$, which fixes the initial longitudinal to transverse pressure ratio $P_L/P_T$. We will consider $\xi_0=[-0.5,\,0,\,10]$ which corresponds, respectively, to a prolate, spherical and oblate distribution in momentum space.
{We will also consider the limit $\xi_0\to \infty$, which means to initialize a system with initial $P_L=0$.}

The goal of this work is to follow the time evolution of the distribution function starting from different sets of initial conditions.
In order to do that we perform a systematic study of the moments of the distribution function given in Eq. \eqref{eq:momentum_moments}.
We now restrict to the case of the Bjorken flow, hence the moments can be rewritten as \cite{Almaalol:2020rnu}:
\begin{align}\label{eq:momentum_moments_BJ}
\mathcal{M}^{mn}(\tau)&=\int dP \, (p\cdot u)^n \, (p\cdot z) ^{2m} \, f(\tau,p)\\&=\int \frac{d^2 p_T dp_w}{(2\pi)^3 p_\tau} p_\tau^n p_w^{2m} \, f(\tau,p)
\end{align}
where the integration measure in the momentum space $dP$ in this case becomes $\dfrac{d^2 p_T dp_w}{(2\pi)^3 p_\tau}$, being
\begin{subequations}\label{eq:strick_moments}
    \begin{gather}
    p_w=p_z\cosh(\eta) - p_0 \sinh(\eta),\\
    p_\tau=p_0\cosh(\eta) - p_z \sinh(\eta),
\end{gather}
\end{subequations}
the momenta associated with the proper time $\tau$ and the space-time rapidity $\eta$, in which $p_z$ is the longitudinal momentum and $p_0$ for an on-shell massless particle corresponds to the energy.
The system flow velocity $u^{\mu}$ and the orthogonal versor $z^{\mu}$ in the case of the Bjorken flow reduce to \cite{Bjorken:1982qr}
\begin{subequations}\label{eq:umu_bjorken}
\begin{gather}
    u^\mu = (\cosh \eta, 0,0,\sinh\eta), \\ z^\mu=(\sinh\eta, 0,0, \cosh\eta).
\end{gather}
\end{subequations}

Some of the previous moments have a clear physical interpretation: $\mathcal{M}^{10}$, $\mathcal{M}^{20}$ and $\mathcal{M}^{01}$ correspond, respectively, to the number density $n$, the energy density $\varepsilon$ and the longitudinal pressure $P_L$. Notice that for a conformal system $\varepsilon=2P_T+P_L$; therefore, we can compute also the transverse pressure from $\mathcal{M}^{20}$ and $\mathcal{M}^{01}$.
The momentum moments encode all the information of the distribution function.
Notice that we will relax the assumption of Bjorken flow in Sec. \ref{sec:no_boostinv}; in that case the expression Eq. \eqref{eq:momentum_moments_BJ} is no more valid and we will use the general definition Eq. \eqref{eq:momentum_moments}.

\begin{figure*}[th!]
    \centering
    \includegraphics[width=\textwidth]{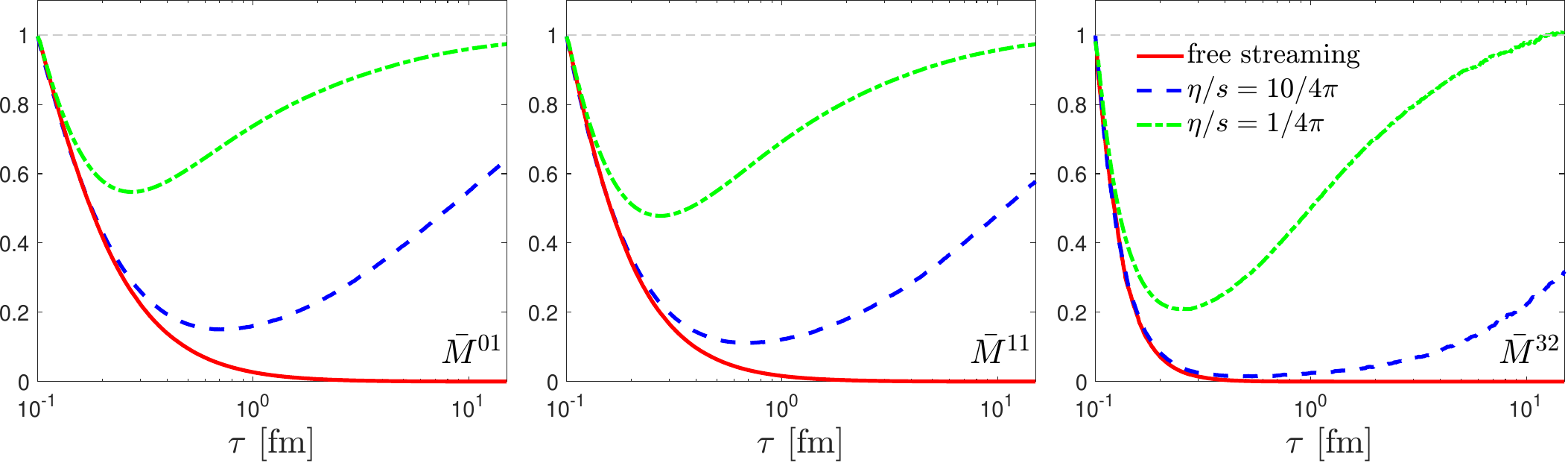}
    \caption{Normalized moments as a function of proper time $\tau$ at mid-rapidity obtained with the RBT approach initialized with $\tau_0= 0.1$ fm, $T_0 = 0.5$ GeV and $\xi_0=0$. Different curves correspond to different values of $\eta/s$: $1/4\pi$ (dot-dashed green), $10/4\pi$ (dashed blue), free streaming $\eta/s \to \infty$ (solid red).}
    \label{fig:freestream}
\end{figure*}

In order to study the approach to the equilibrium we make use of the normalized moments $\overline M^{mn}=\mathcal{M}^{mn}/\mathcal{M}^{mn}_{eq}$: notice that in the limit of isotropic and thermal equilibrium $\overline M^{mn}\to 1$.
In Fig. \ref{fig:freestream}, we show the results of the RBT approach for some of the normalized momentum moments of the distribution function as a function of the proper time at midrapidity $|\eta|<0.02$. The calculation starts at initial proper time $\tau_0= 0.1$ fm, with initial temperature $T_0 = 0.5$ GeV and without initial anisotropy ($\xi=0$, $\alpha=1$). We consider two different values of the viscosity over entropy density ratio: $4\pi\eta/s=1$ and $4\pi\eta/s=10$. We observe that all moments reach the isotropic and thermal limit for large times. In particular, in the case of smaller $\eta/s$ the equilibrium is reached earlier with respect to the case with larger specific viscosity: for all the moments presented $\overline M^{mn}\to 1$ at a time $\tau\sim2$ fm for $4\pi\eta/s=1$, while that limit is reached at $\tau>20$ fm for $4\pi\eta/s=10$.
This is not surprising, since the scattering rate is inversely proportional to the specific viscosity.
Momentum moments with different $m,n$ probe the behavior of the distribution function in different regions in $p$ and $p_z$ of the phase-space.
Due to the chosen initial condition ($P_L/P_T=1$), the moments start from $\overline M^{mn}=1$, deviate from the equilibrium limit and then approach it again, with different time scales depending on the specific viscosity of the system, as mentioned before. However, the amount of deviation from the equilibrium during the evolution depends on the powers of $p_z$ and $p$. Moving from the left to right in the plot, we probe higher powers of $p_z$ and $p$. We observe that for higher order $m,n$ there is a larger deviation of the normalized moments from the equilibrium limit.
This is easily understood since high-energy particles contributing more to the high order moments are expected to thermalize later, as they need a larger number of scatterings in order to equilibrate with the surrounding medium.
In Fig. \ref{fig:freestream} the curves obtained with finite specific viscosity are compared with the case of free-streaming, that corresponds to $\eta/s\to\infty$. In our code this limit is achieved imposing that the scattering cross section is zero.
We observe that, even though the free-streaming lines start from the equilibrium limit $\overline M^{mn}=1$, they quickly deviate from it due to the longitudinal expansion and, since there are no microscopic mechanisms that tend to recover the local equilibrium, the moments move away indefinitely from the equilibrium limit.
We notice that the curves at finite viscosity follow the free-steaming case at initial time and then deviate from it approaching again the equilibrium limit. This is due to the competition between two different effects: the longitudinal expansion and the collisions between particles.
The momentum exchange due to collisions tends to isotropize the system, producing the effect of reducing the strong anisotropy along the longitudinal direction and drives the normalized moments toward the isotropic limit $\overline M^{mn}\to 1$.
Since lower viscosities correspond to larger scattering rate, the system tends to isotropize faster in the case $4\pi\eta/s=1$ with respect to the higher viscosity case $4\pi\eta/s=10$.
However, at initial times, the longitudinal expansion is so violent that its effect dominates over the collisions. 
Therefore, the appearance of a minimum in the curves $\overline M^{mn}$ with $n>0$ and the subsequent increase toward the isotropic limit is connected to the dominance of collisions. 

\begin{figure*}[th!]
    \centering
    \includegraphics[width=\textwidth]{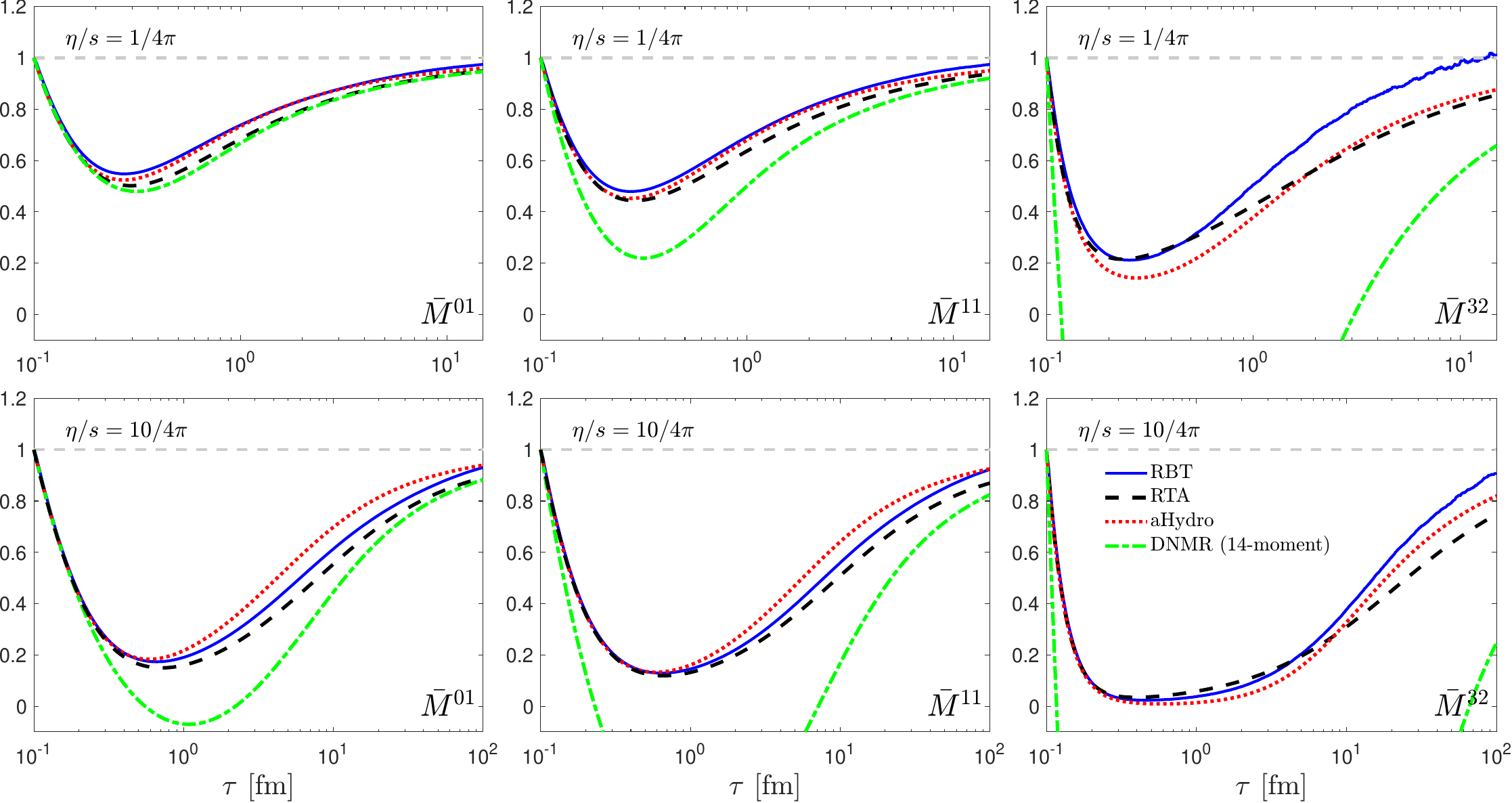}
    \caption{Normalized moments as a function of proper time $\tau$ at mid-rapidity, for $\tau_0= 0.1$ fm, $T_0 = 0.5$ GeV and $\xi_0=0$. Different curves correspond to different models: RBT (solid blue), Boltzmann RTA (dashed black), anisotropic hydrodynamics (dotted red) and DNMR with 14-moment approximation (dot-dashed green). The upper panels correspond to the case with $\eta/s=1/4\pi$, while the lower ones to $\eta/s=10/4\pi$.}
    \label{fig:Mmn_approaches}
\end{figure*}

In Fig. \ref{fig:Mmn_approaches} we show, for the same initial conditions explained in Fig. \ref{fig:freestream} and the two values of specific viscosity $4\pi\eta/s=1$ and $4\pi\eta/s=10$, the normalized moments $\overline M^{01},\overline M^{31}, \overline M^{32}$ obtained with the RBT approach in comparison to the results of different approaches.
The solid blue curve is the result of the simulation obtained with RBT, while the dashed black, the dotted red and the dot-dashed green lines are the results, respectively, of the RTA Boltzmann equation, anisotropic hydrodynamics and DNMR viscous hydrodynamics in the 14-moment approximation.

With the RBT approach, which includes a full collision integral that is gauged to a fixed viscosity over entropy density ratio, one can go from free streaming ($\eta/s\gg1$) to a strongly interacting medium (small $\eta/s$).
The agreement between full RBT and RTA depends on both the viscosity and the order of the moments. Indeed, for larger viscosity we get a better agreement between the two approaches for all the three moments, but for smaller viscosity the discrepancy increases, especially for $\overline M^{32}$.
A similar trend is present for the difference between aHydro and the full Boltzmann approach. {However, it is interesting to notice that, especially for higher order moments, the equilibration is faster for RBT than for ahydro and RTA. As far as DNMR is concerned, it is not surprising that the agreement is better for lower order moments and breaks down for higher ones, since DNMR is constructed by taking just the first order moments of the Boltzmann distribution function, and therefore turns even negative for higher order. It would be interesting, however, to compare our results with more recent formulations of DNMR, in which one goes beyond the 14-moment approximation \cite{Denicol:2012cn, Wagner:2023joq}. }\\
In order to emphasize the universal behavior of attractors, it is of common use in literature to plot the system evolution, for instance the moments of the distribution function, with respect to the scaled time $\tau/\tau_{eq}$ \cite{Jankowski:2023fdz, Kurkela:2018vqr, Giacalone:2019ldn}. In RTA and hydrodynamics an expression for the relaxation time $\tau_{eq}$ is directly needed for their formulation because it governs the evolution equations. It is often chosen $\tau_{eq}^{RTA}=5\, (\eta/s)\,/T$ \cite{Huovinen:2008te, Betz:2010cx, Denicol:2010xn, Denicol:2011fa}.
In the framework of our full Boltzmann approach it is possible to recover the universal behavior of attractors dividing the proper time by a timescale naturally emerging in our kinetic approach that is related to the collisional dynamics. More precisely, the attractor behavior is manifest when the proper time is scaled by  the average collision time per particle $\tau_{coll}$:
\begin{equation}\label{eq:taucoll}
    \tau_{coll}= \frac 12 \tonde{ \frac{1}{N_{\text{part}}} \frac{\Delta N_{\text{coll}} }{\Delta t} }^{-1}
\end{equation}
where $N_{\text{coll}}$ is the number of collisions occurring in a certain time interval $\Delta t$ in a given volume with $N_{\text{part}}$ test particles, and {the factor $1/2$ accounts for double counting}.
The quantity $\tau_{coll}$ determines how quickly the system approaches the equilibrium. The system equilibration, within a kinetic approach, is connected to the particle collisions, which have the role to isotropize and thermalize the system, inducing a loss of information about the initial conditions.\\

More precisely, what is relevant is defined as the transport relaxation time, that for a system of massless particles interacting through an isotropic cross-section, is \cite{Plumari:2012ep}: 
\begin{equation}
    \tau_{tr} = \frac{1}{n\, \sigma_{tr}} = \dfrac{1}{\frac{2}{3}n\, \sigma} = \frac 32 \tau_{coll}
\end{equation}
Hence we can naturally define the RBT relaxation time:
\begin{equation}\label{eq:tau_RBT}
    \tau_{eq}^{\text{RBT}} = \tau_{tr} = \frac 32 \tau_{coll}.
\end{equation}
For a system in chemical equilibrium the relation between entropy density and particle density is given by $s=4 n$. From the Chapman-Enskog expansion $\eta = 1.2\, T/\sigma$, therefore one finds $\tau_{eq}^{RBT}=\tau_{tr}=\tau_{eq}^{RTA}$.\\
In our case, as anticipated in Sec. \ref{sec:RBT}, we have to take into account the fugacity $\Gamma$ of the system, due to the particle number conservation. Therefore, the relation between entropy density and particle density becomes $s = n\, (4-\log \Gamma)$. This means that, in order to have a consistent comparison, $\tau_{eq}^{RTA}$ must be corrected by a factor $(1 - \log \Gamma /4)$.

\begin{figure}[t!]
    \centering
    \includegraphics[width=\columnwidth]{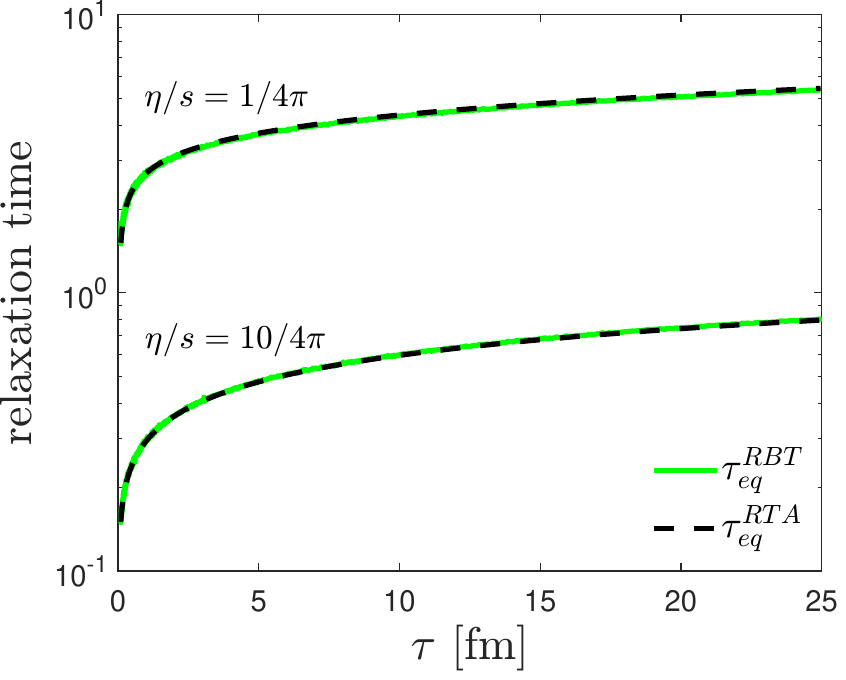}
    \caption{Comparison for two different values of specific viscosity between the relaxation time $\tau_{eq}^{\text{RBT}}$ (solid green line) according to Eq. \eqref{eq:tau_RBT} and $\tau_{eq}^{\text{RTA}}$ (dashed black line) in Eq. \eqref{eq:tau_RTA} where $T(\tau)$ is the one calculated in the simulation. The initial conditions are the same used in Fig. \ref{fig:freestream}.}
    \label{fig:tau_eq}
\end{figure}
\begin{figure}[t!]
    \centering
    \includegraphics[width=\columnwidth]{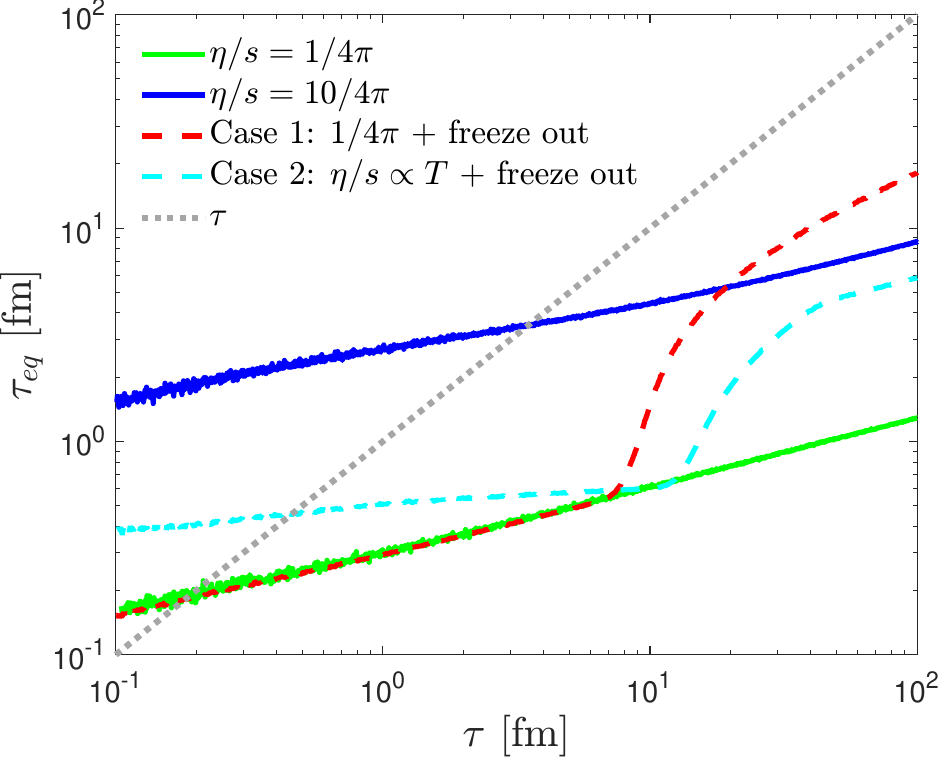}
    \caption{Comparison among relaxation times  $\tau_{eq}^{\text{RBT}}$ in the RBT approach for different values of $\eta/s$: $1/4\pi$ (green line), $10/4\pi$ (blue line), and, referring to \ref{sec:Tdep_visco} Case 1 (red dashed line) and Case 2 (cyan dashed line).}
    \label{fig:tau_eq_RBT}
\end{figure}

In Fig. \ref{fig:tau_eq}, we show the comparison of the $\tau_{eq}$ adopted in RTA and hydrodynamics with the one used in our approach, suitably modified to remove the effect of fugacity. There is a full agreement during the whole time evolution for the range of specific viscosity explored in this paper, as evident from the lines with $4\pi\eta/s=1$ and $4\pi\eta/s=10$ shown in the plot.
{In Fig. \ref{fig:tau_eq_RBT} we plot $\tau_{eq}^{RBT}$ with respect to $\tau$ for $4\pi\eta/s=\{1,10\}$ and for $\eta/s(T)$ as parametrized in Sec. \ref{sec:Tdep_visco}, as well as the bisector of the $\tau$-$\tau_{eq}$ plane. By comparing this plot and Fig. \ref{fig:freestream} one can see that when the collisions start to dominate ($\tau\lesssim \tau_{eq}$), the curves pass by the minima. This is quite patent by looking at the $\eta/s=1/4\pi$ case, while the case with $\eta/s = 10/4\pi$ is slightly more difficult to be read, since the position of the minimum and even the extension of the free-streaming-dominated region strongly depends on the moment's order.} 

In the following sections we will show the approach to the attractor behavior in our framework by showing the moments of the distribution function versus the proper time scaled by the equilibration time in Eq. \eqref{eq:tau_RBT}.

\section{Pull-back and forward attractors}
\label{sec:pullback_forward}

It has been shown in the literature, within the framework of relativistic hydrodynamics formulations \cite{Strickland:2017kux, Alalawi:2020zbx, Heller:2020anv, Heller:2015dha, Chattopadhyay:2019jqj, Jaiswal:2019cju}  as well as microscopic models based on relativistic kinetic theory (EKT \cite{Kurkela:2018vqr, Kurkela:2019set, Almaalol:2020rnu}, RTA \cite{Strickland:2018ayk, Heller:2018qvh, Blaizot:2017ucy, Kamata:2020mka}, BAMPS \cite{Ambrus:2021sjg}, classical Yang-Mills equations \cite{Berges:2013eia, Berges:2013fga} and AdS-CFT \cite{Heller:2011ju,Romatschke:2017vte, Kurkela:2019set}), that an universal behavior emerges during the system evolution in some dynamical quantities, such as the longitudinal pressure normalized to its equilibrium value or the ratio between longitudinal and transverse pressures, suggesting a loss of information about the initial conditions. This is the so called attractor behavior.
Such scaling is manifest when the proper time is divided with the relaxation time $\tau_{eq}=5(\eta/s)/T(\tau)$.

Following \cite{Heller:2020anv,Alalawi:2022pmg}, we look for both forward (or late-time) and pull-back (or early-time) attractors. As mentioned in the introduction, by forward attractor we mean the late-time convergence of systems initialized with the same $\tau_0, T_0$ and $\eta/s$ but with different initial anisotropies $\xi_0$; by pull-back attractor we mean the convergence of curves starting from different initial scaled time $\tau_0/\tau_{eq,0}$, i.e. varying $\tau_0$, $T_0$ or $\eta/s$. For the sake of simplicity, we will keep $T_0$ fixed and vary the ratio $(\eta/s)/\tau_0$. We will see that it is only the ratio $\tau_0/\tau_{eq,0}$ which matters when plotting curves as functions of the scaled time.
{In viscous and anisotropic hydrodynamics it is clearly visible when the system of ODEs is written as a single equation for $\varphi = \varphi(\pi/\varepsilon) = 2/3 + \pi/(3\varepsilon)$ as a function of $w = \tau/\tau_{eq}$ \cite{Strickland:2018ayk}. Eventually, the normalized moments in Eqs. (\ref{eq:vhydro_moments}, \ref{eq:ahydro_moments}) depend only on the function $\pi/\varepsilon (w)$, which is dependent only on the initial value $w_0 = \tau_0 T_0 /(\eta/s)$. Thus, if one changes two or more of these parameters without changing $w_0$, this will not affect the solutions $\overline M^{nm}(w)$. }\\
The universal behavior of various systems in the beginning of the evolution, due to the dominant longitudinal expansion (similar to a free streaming), as explained in the previous section, can be emphasized by looking at pull-back attractors. 
Regardless the $\eta/s$, i.e. the scattering rate of the medium, all systems share the same initial evolution pattern: this can be easily visualized by plotting moments with respect to the scaled proper time $\tau/\tau_{eq}$ for different systems with same initial conditions (temperature and anisotropy) but different $\tau_0/ (\eta/s)$, which means changing the initial scaled time.

\begin{figure*}[th!]
    \centering    \includegraphics[width=\textwidth]{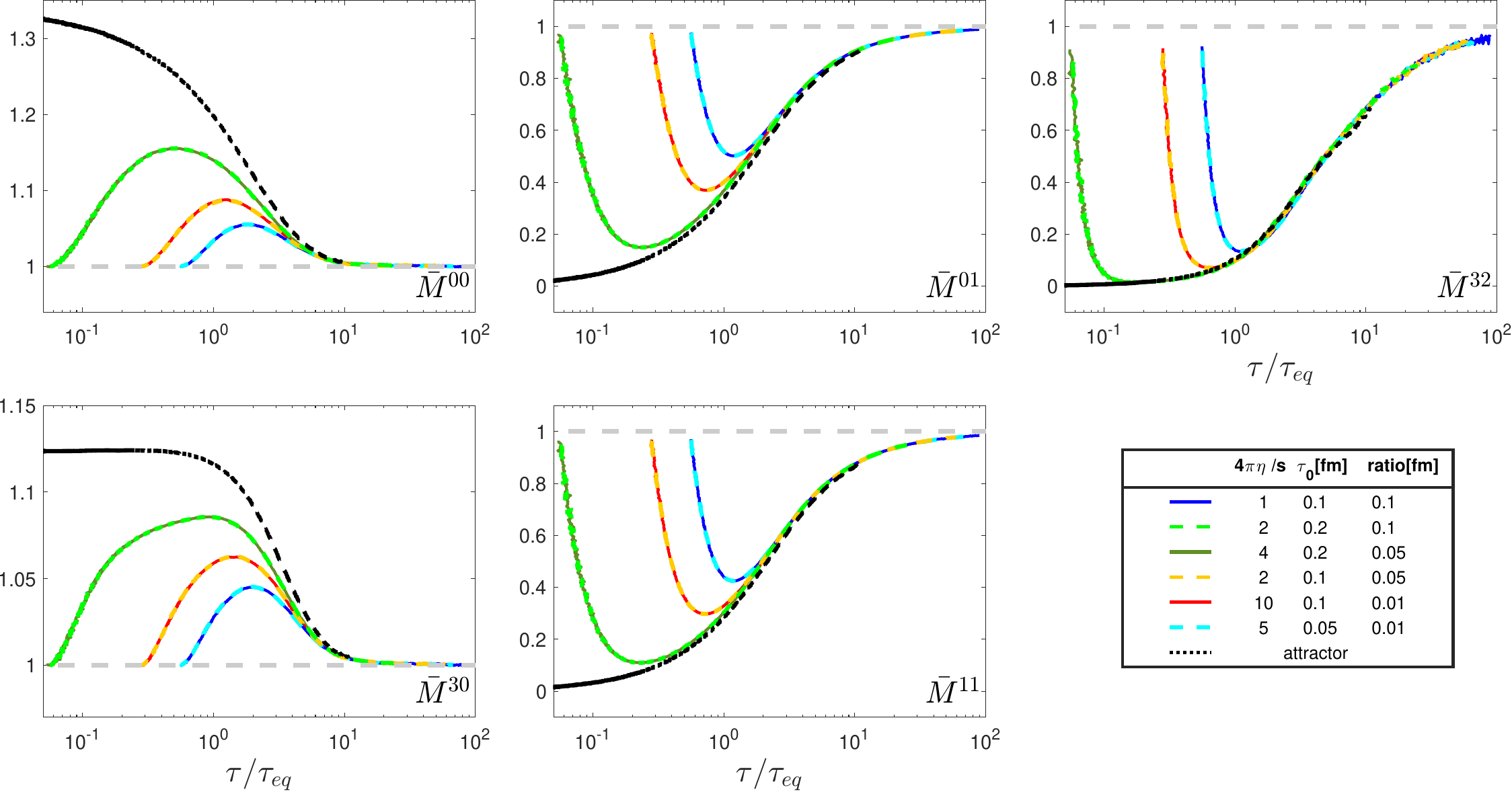}
    \caption{Normalized moments as a function of $\tau/\tau_{eq}$ within the RBT approach. Different curves correspond to three different values of $\tau_0/(\eta/s) = [0.1,\,0.05,\,0.01]$ fm, highlighting the pull-back attractor behavior. The black dotted line is the attractor curve in RBT as described in the text. The initial temperature $T_0 = 0.5$ GeV is fixed.}
    \label{fig:pullback}
\end{figure*}

In Fig. \ref{fig:pullback} we show the momentum moments of the distribution function as a function of the scaled time $\tau/\tau_{eq}$ obtained with the full Boltzmann approach starting with initial temperature $T_0=0.5$ GeV and initial zero anisotropy ($\xi_0=0$). The different curves are obtained by changing the value of $\tau_0$ and $\eta/s$. We notice, however, the simulations with different $\tau_0$ and $\eta/s$ but same ratio $\tau_0/(\eta/s)$ give curves which lie on top of each other, as visible in the figure by looking for example at the {solid light green} and {dashed dark green} lines with have both $\tau_0/(\eta/s)=0.01$ (and similarly for the other two cases). 

Curves with different values of $\tau_0/(\eta/s)$ start from different $\tau/\tau_{eq}$ values. The almost free initial expansion drives the system away from equilibrium until the collisions start to dominate: here the normalized moments reach a minimum/maximum (depending on the order $n$ of the moment) and immediately afterwards they reach the attractor (dotted black line), all lying on the same curve.
Therefore, the universal scaling seen in other models (RTA, EKT, AdS/CFT\dots) is present also in simulations with the relativistic Boltzmann approach with full collision integral.

\begin{figure*}[th!]
    \centering
    \includegraphics[width=\textwidth]{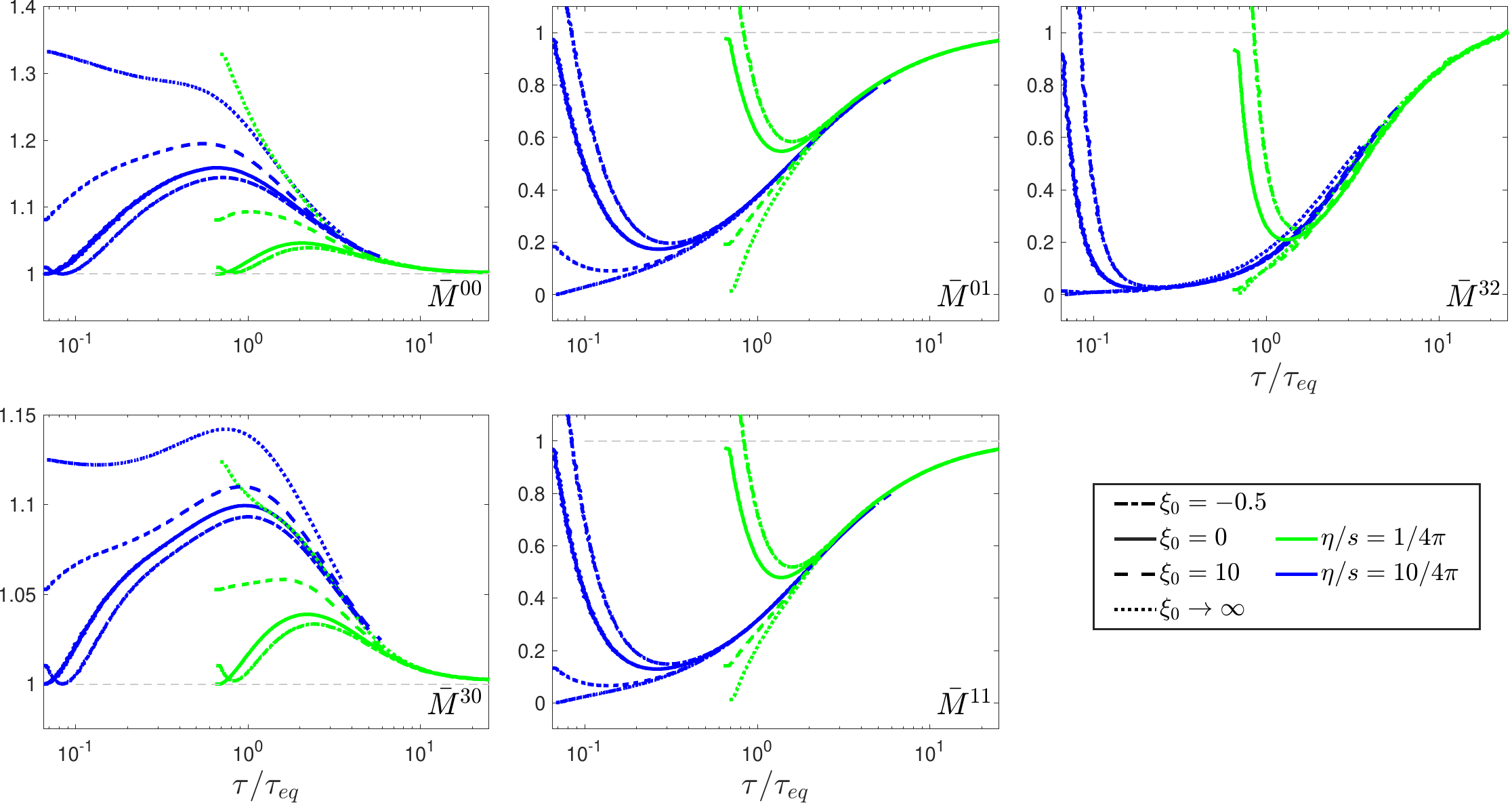}
    \caption{Normalized moments as a function of $\tau/\tau_{eq}$ within the RBT approach. Different curves correspond to four different values of initial anisotropy $\xi_0 = [-0.5,\,0,\,10,\, +\infty]$, highlighting the forward attractor behavior. The two colors correspond to two different values of $\eta/s$. We fix $T_0= 0.5$ GeV and $\tau_0 = 0.1$ fm.}
    \label{fig:forward}
\end{figure*}

\begin{figure*}[th!]
    \centering
    \includegraphics[width=\textwidth]{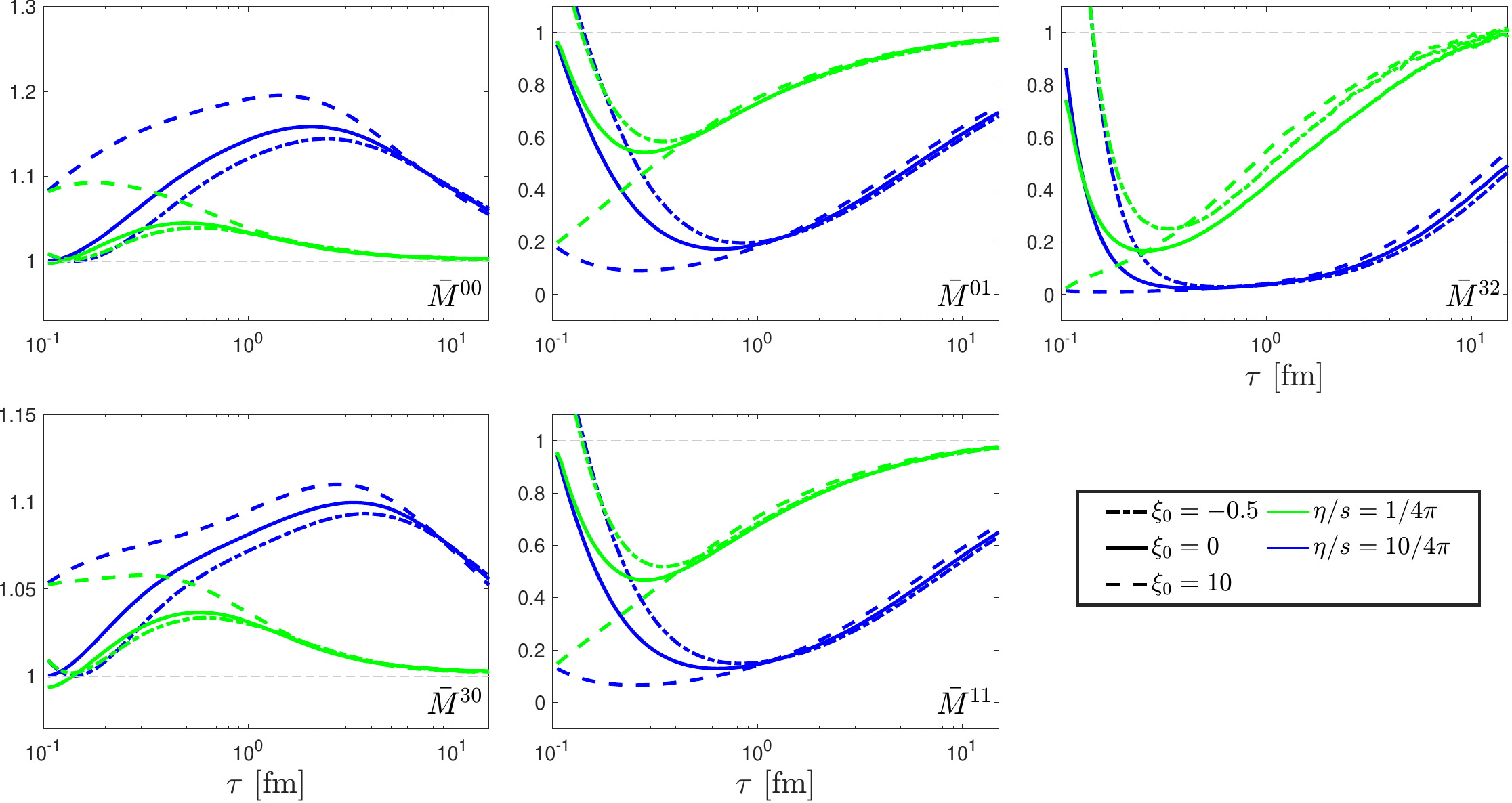}
    \caption{Normalized moments as a function of $\tau$ within the RBT approach. Colors and linestyles are the same of Fig \ref{fig:forward}.}
    \label{fig:forward_tau}
\end{figure*}

In Fig. \ref{fig:forward} we show moments as a function of scaled time at midrapidity. The results have been obtained with the full relativistic Boltzmann approach for two values of $\eta/s$ and four different values of the anisotropy parameter $\xi_0$, while keeping fixed the initial time and temperature.
Similarly to what has been seen in other models, we notice that the larger the specific viscosity is, the smaller is the scaled time needed to reach the attractor curve: for $\eta/s=1/4\pi$ ({green lines}) the simulations converge at $\tau_{attr} \approx 1.5\,\tau_{eq}$, whereas for $\eta/s =10/4\pi$ ({blue lines}) the attractor is reached at about $\tau_{attr} \approx 0.2\, \tau_{eq}$. \\
Notice that when we plot the moments as a function of proper time (instead of the time scaled with the equilibration time) as in Fig. \ref{fig:forward_tau}, they converge toward similar values and then evolve in a similar way, although the single attractor curve is not clearly visible. In this case, it is clear that the convergence of the simulations with different anisotropies is reached earlier for smaller specific viscosity, because a smaller $\eta/s$ corresponds to a larger scattering rate, leading to a quicker loss of memory of the initial condition details. However, it is interesting to notice that even though there's a factor 10 between the two different specific viscosities $\eta/s$, there's a much smaller factor between the times in which attractors are reached. This suggests that the mechanism which brings the system to the attractor is not due only to the collisions, but there is also a strong contribution by the initial longitudinal expansion.}

From Fig. \ref{fig:forward}, we observe that for smaller specific viscosity when the attractor is reached the normalized moments are closer to 1 with respect to the case with larger $\eta/s$. For example $\overline{M}^{01}=\overline{P}_L=0.6$ for $\eta/s=1/4\pi$ and $\overline{M}^{01}=\overline{P}_L=0.23$ for $\eta/s=10/4\pi$.
This is due to the fact that for larger specific viscosity the effect of the strong longitudinal expansion dominates for longer time, so that the system has reached a higher degree of anisotropy when it deviates from the free-streaming trend (see Fig. \ref{fig:freestream}) and thanks to the collisions approaches the universal behavior. 

\begin{figure*}[th!]
    \centering
    \includegraphics[width=\textwidth]{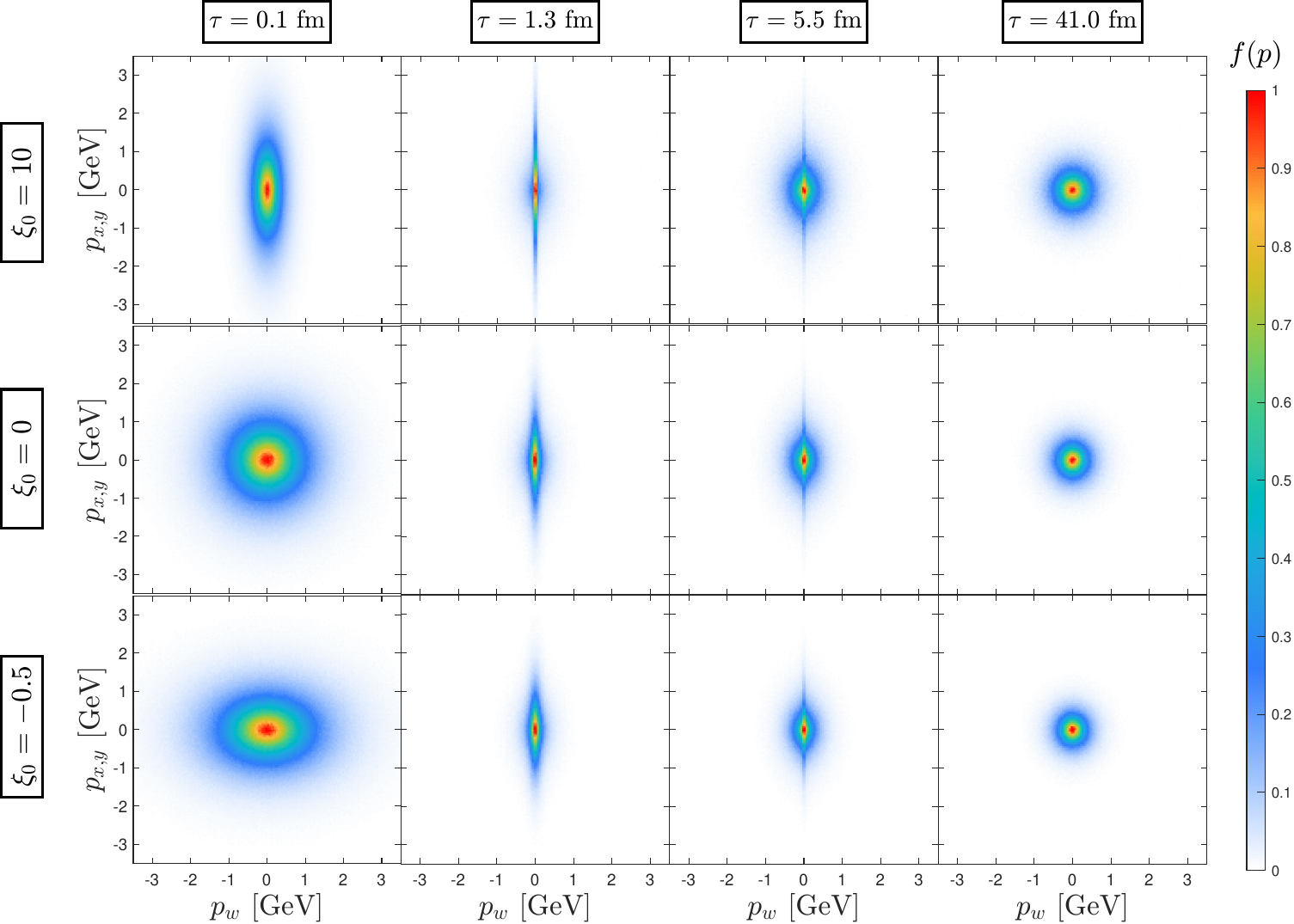}
    \caption{Contour plots of the normalized distribution function in the ($p_{x,y},p_w$) space for RBT computed at mid-rapidity. Different columns refer to different $\tau$ values; different rows to different initial anisotropies. We fix $T_0 = 0.5 $ GeV, $\tau_0= 0.1$ fm and $\eta/s = 10/4\pi$. }
    \label{fig:distr_contour}
\end{figure*}

In order to further elucidate this point, we show in Fig. \ref{fig:distr_contour} the time evolution of the profile of the distribution function on the $p_w$-$p_T$ plane for {$\eta/s=10/4\pi$} and three different values of the initial anisotropy parameter: $\xi_0=[-0.5,\,0,\,10]$. We observe that, although the shape in momentum space is very different at the initial time of the simulation $\tau_0=0.1$ fm$/c$, after about {1.3 fm$/c$} the distributions become very similar and in all three cases elongated more along the $p_T$ than along the $p_w$ axis. This convergence of the anisotropic shape of the distribution function is driven by both the strong initial longitudinal expansion and the collisions. After this convergence the profiles continue to evolve in a similar way toward the isotropic limit.
The profile of the distribution function has been studied in Ref. \cite{Strickland:2018ayk}, pointing out that the distribution function contains two components: an anisotropic part which becomes more and more squeezed in longitudinal momentum as time evolves; a more isotropic piece which dominates at late times.
A similar distinction is clearly visible also in our simulations, especially by looking at the third column ($\tau = 5.5$ fm).
{This  two-component shape can account for the quite peculiar behavior of the normalized moments $\overline M^{n0}$ in Fig. \ref{fig:pullback} and \ref{fig:forward} ($\overline M^{10}$ and $\overline M^{30}$ are identically 1 due to the matching conditions). It is manifest that the attractor is reached quite later in both plots. As highlighted in \cite{Strickland:2018ayk} this is due to the fact that these moments are strongly sensitive to the region of the phase space where $p_w \sim 0$, which is exactly that populated by the squeezed component of the distribution function. As a consequence, these moments are forced to reach the attractor and to thermalize later.}
\begin{figure}[th!]
    \centering
    \includegraphics[width=\columnwidth]{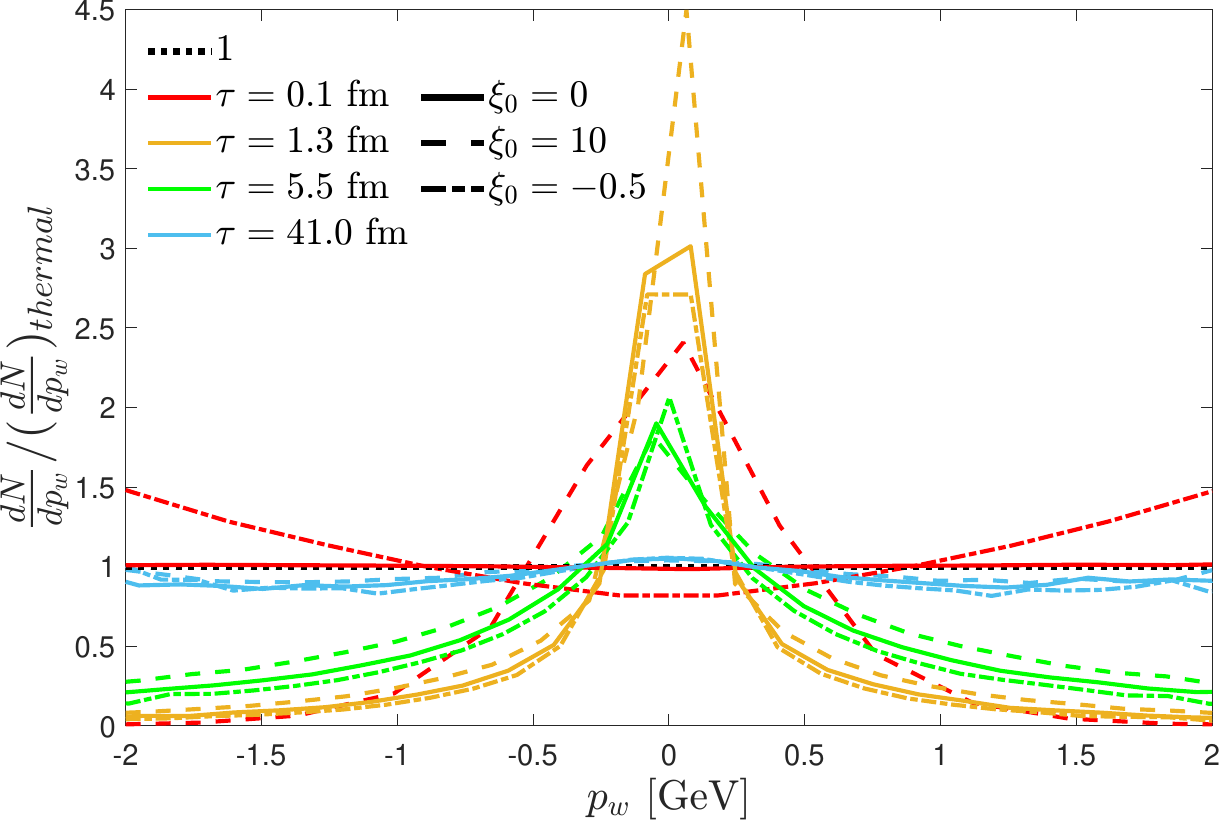} \\
    
    \includegraphics[width=\columnwidth]{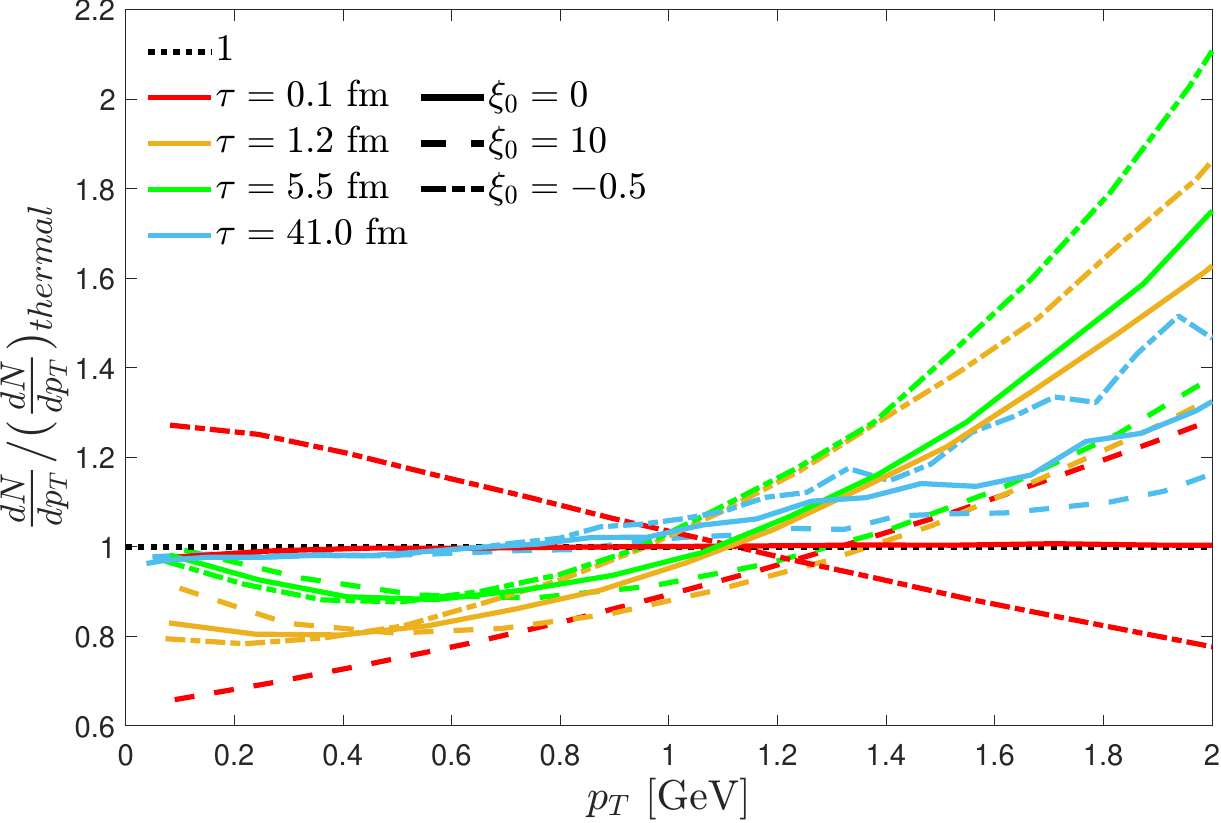}
    \caption{Upper panel: ratio between the computed $dN/dp_w$ and the thermal one corresponding to the effective temperature $T$. Lower panel: ratio between the computed $dN/dp_T$ and the thermal one. In both panels, different colors correspond to the same proper times shown in Fig. \ref{fig:distr_contour} and different styles to the same $\xi_0=[0,\,10,\,0.5]$ respectively for solid, dashed and dot-dashed lines. The initial conditions are those of Fig. \ref{fig:distr_contour}. }
    \label{fig:ratio_spectrum}
\end{figure}

While Fig. \ref{fig:distr_contour} gives us information about the isotropization of the system, in order to investigate its thermalization, we show in Fig. \ref{fig:ratio_spectrum} the time evolution of the ratio of the phase-space distribution function over the thermal Boltzmann distribution corresponding to the effective temperature of the system at the considered time. In particular, the top panel presents the distribution in longitudinal momentum {$dN/dp_w$} integrated over $p_T$ and the bottom panel is the distribution in transverse momentum {$dN/dp_T$} integrated over $p_w$, both computed at midrapidity {$|\eta|<0.75$}.
From the top plot, we see that at initial time $\tau_0=0.1$ fm$/c$ the lines corresponding to different initial anisotropies are very different between each other. In particular, $\xi_0=0$ corresponds initially to a thermal distribution by construction. We observe that after {1.0 fm$/c$}, that corresponds roughly to the time in which the system reaches the attractor, due to the collisional dynamics for $\xi_0=0$ the spectrum become more populated at high longitudinal momentum with respect to the thermal distribution, showing large deviations from 1. The other two curves $\xi_0=-0.5$ and $\xi_0=10$ are from the beginning far from equilibrium. As time evolves, the three simulations get closer to each other, suggesting the universal behavior in the longitudinal momentum component of the phase-space distribution function, and show a similar trend in approaching the equilibrium limit 1.
From the bottom panel, we observe a similar behavior in approaching the thermal distribution, as observed in $dN/dp_w$. The collisional dynamics affects the transverse momentum spectrum (both thermal and non-thermal) by moving particles from high to low $p_T$. The rate of this shift in $p_T$ depends on the initial anisotropy of the distribution. From the plot we notice that in the case $\xi_0=10$ the shift of particles from high to low $p_T$ happens faster than for the other two cases. Moreover, the approach toward equilibrium is slower for particles with high $p_T$.
\begin{figure*}[th!]
    \centering
    \includegraphics[width=\textwidth]{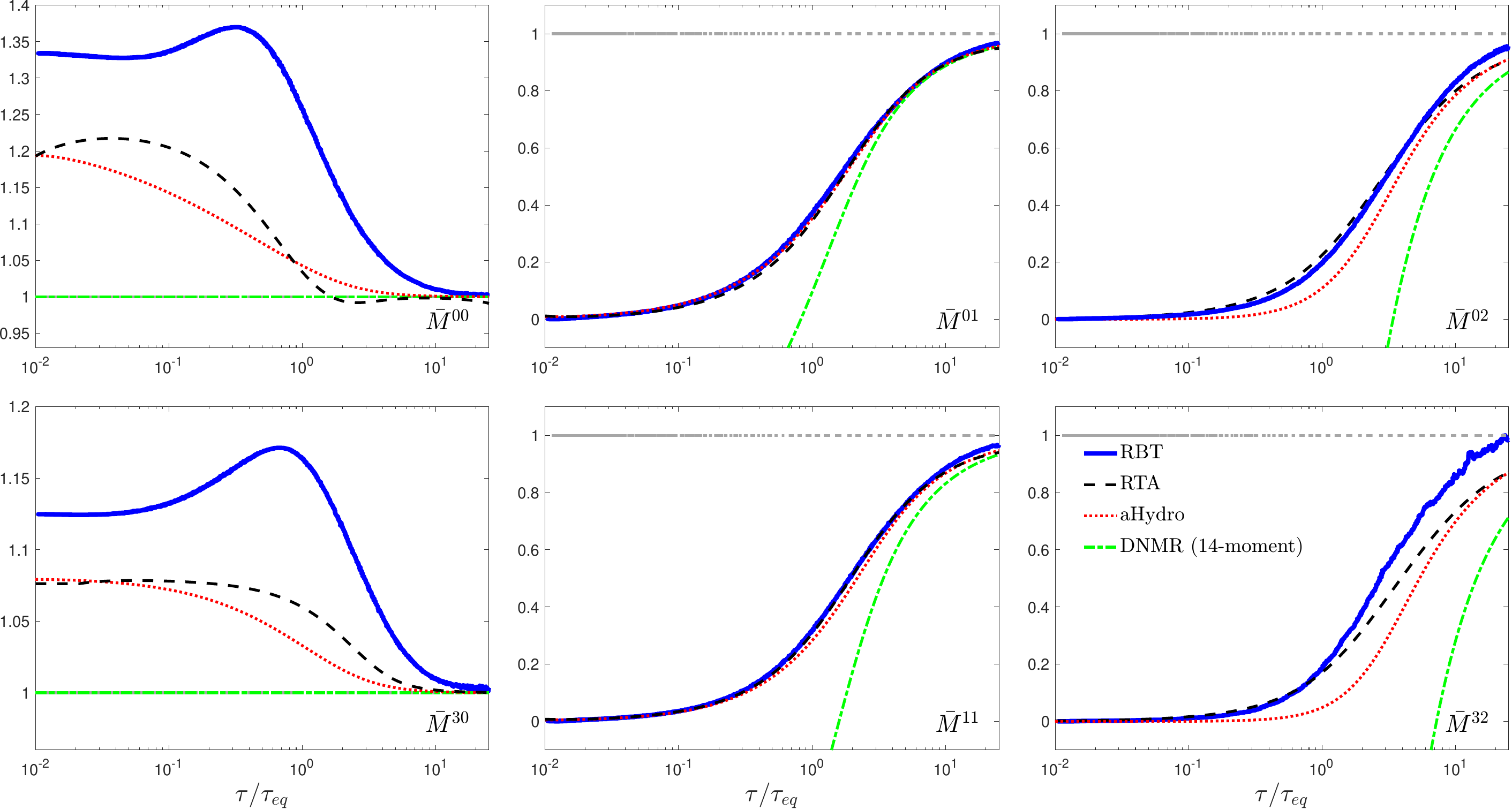}
    \caption{Attractor curves as function of the scaled time $\tau/\tau_{eq}$ for different models as described in the text: RBT (solid blue), RTA (dashed black), aHydro (dotted red) and DNMR with the 14-moment approximation (dot-dashed green).}
    \label{fig:attractor_models}
\end{figure*}

In Fig. \ref{fig:attractor_models} we show the comparison of the attractor curves obtained with our Full Boltzmann approach ({solid blue line}) with those found in RTA ({dashed black line}), aHydro ({dotted red line}) and DNMR theory ({dot-dashed green line}). In our simulation the prescription used to obtain the attractor curve is the one used in several approaches \cite{Blaizot:2017ucy, Romatschke:2017vte,Heller:2015dha}, which is equivalent to initialize the distribution function with an initial zero longitudinal pressure and take the limit $\tau_0/(\eta/s)\to 0$. In our RBT approach the case $P_L=0$ is obtained by imposing the condition $Y=\eta$ (momentum rapidity equal to space-time rapidity), which implies $p_w=0$ for every test particle and, consequently, $\mathcal M^{nm}=0$ with $m>0$ and therefore also $P_L=0$. 
In the RTA case, the attractor curve is obtained by taking as the initial condition $\xi_0 \to \infty$ and the initial proper time $\tau_0\to 0$ \cite{Strickland:2018ayk}. These two conditions are equivalent to those we impose: the condition $\xi\to \infty$ forces us to have $p_w=0$ in the distribution function and therefore $P_L=0$ as an initial condition; concerning the second condition, in our prescription we require $\tau_0/(\eta/s)\to 0$, as we have already mentioned that the ratio is the important quantity and not only $\tau_0$. 
{In the hydrodynamic calculations \cite{Strickland:2017kux} the attractor curve is obtained by taking the limit $w\to 0$ for the previously defined function $\varphi(w)= 2/3 + 1/3\,\pi/\varepsilon (w)$ and imposing $\varphi$ to be finite, where $w=\tau/\tau_{eq}$ and $\varphi$ can be written in terms of the longitudinal to transverse pressure ratio by $P_L/P_T=(3-4\varphi)/(2\varphi -1)$.
In aHydro, the condition is fulfilled when $\varphi=3/4$, which implies $P_L=0$ \cite{Strickland:2017kux}}.
We observe that the full Boltzmann, RTA and aHydro approaches give similar results for moments with one power of $p_w$; at higher momentum moments we see a deviation between {RBT} and RTA which gets larger with the increasing order, up to 10-15\% for $\overline M^{32}$  for $\tau/\tau_{eq}>10$. The discrepancy is still more important when comparing {RBT} and aHydro for $\tau/\tau_{eq}<10$, that is also a region of disagreement between aHydro and RTA. The discrepancy is quite significant for higher order moments where however the RBT approach is parametrically more appropriate than hydrodynamics. For instance in RBT $\overline M^{32}$ reaches the near equilibrium value of 0.8 at $\tau/\tau_{eq}\sim 5$, while in aHydro this occurs only at $\tau/\tau_{eq} \sim 10$.

For completeness we added also the curves corresponding to calculations with DNMR theory.
We see all the normalized moments in the DNMR theory, as well as in the other viscous hydrodynamics models, become negative in a certain range, therefore losing their physical meaning. This is due to the fact that in the initial stage the system is dominated by the free-streaming which drives it in a regime where the near-equilibrium assumption of DNMR is no longer valid {\cite{Strickland:2018ayk}}. {Furthermore, the $\overline M^{n0}$ in DNMR are identically 1 for construction. As far as these modes are concerned, it is quite relevant that they show the strongest disagreement between the different frameworks. As outlined before, these moments explore the $p_w\sim0$ region of the phase space (the squeezed component of Fig. \ref{fig:distr_contour}) which is heavily sensitive to the free streaming expansion. Differently than $\overline{M}^{n0}$ with $n>0$, these moments thermalize later in RBT than in other RTA or aHydro and even converge to quite different attractor curves. This suggests that, although the isotropic component of the distribution function is described by these approaches similarly than in RBT, the same cannot be said about the squeezed one, which is the most far-from-equilibrium component of the evolving distribution function.} 
{In order to account for this two-component picture an extension of aHydro has been introduced in \cite{Alalawi:2020zbx}, in which the distribution function ansatz explicitly has a free-streaming and an equilibrating component, as we report in Fig. (\ref{fig:distr_contour}). This of course allows a better agreement with RTA especially for $\overline M^{n0}$.}\\

\section{Temperature dependent specific viscosity}\label{sec:Tdep_visco}

The studies on attractors in the current literature have been performed assuming a constant specific viscosity $\eta/s$.
However, we know from lattice QCD calculations and phenomenological models that $\eta/s$ should depend on temperature and is expected to have a minimum close to the critical temperature \cite{Csernai:2006zz, Plumari:2013bga, Yang:2022ixy}.

\begin{figure}[h]
    \centering
    \includegraphics[width=\columnwidth]{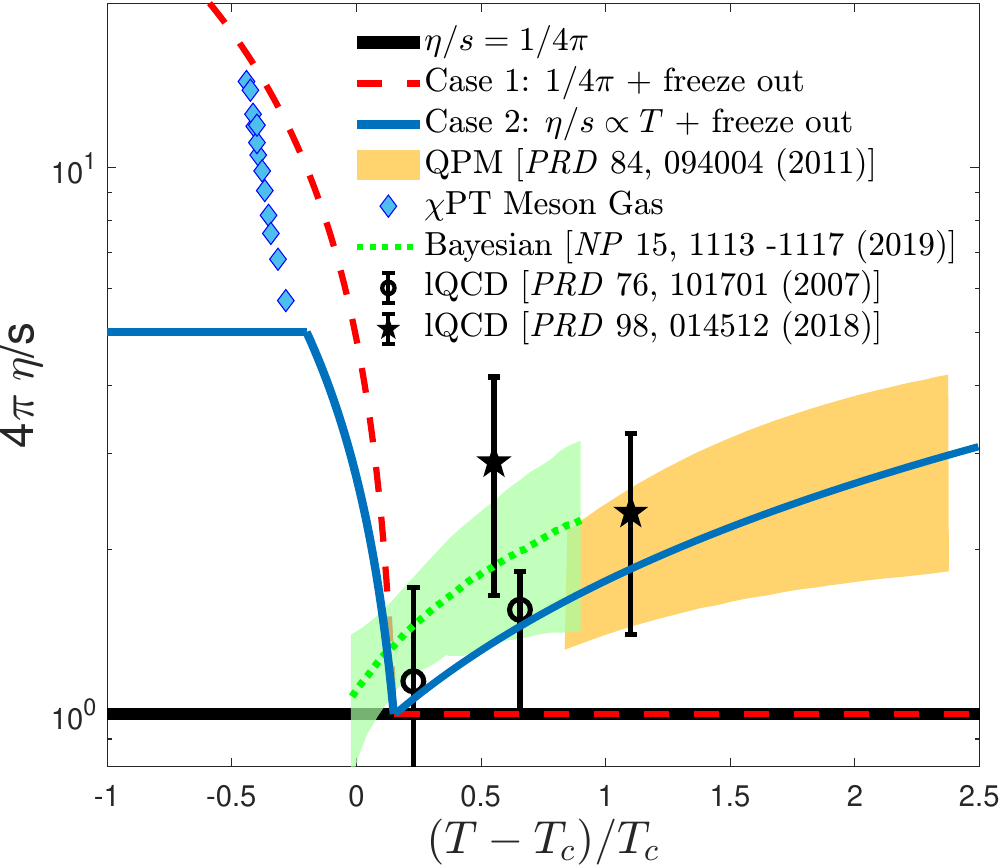}
    \caption{Different temperature-dependent $\eta/s$ parametrizations. Constant $\eta/s=1/4\pi$ (solid black), Case 1 (red dashed) and Case 2 (blue solid) are the ones used in our simulations. Case 1 corresponds to constant $\eta/s=1/4\pi$ for higher temperature and linearly rising at lower temperature, to simulate an almost sudden freeze-out; Case 2 is a more realistic parametrization with a minimum close to $T_C$.}
    \label{fig:etas_temp}
\end{figure}
\begin{figure*}[th!]
    \centering
    \includegraphics[width=\textwidth]{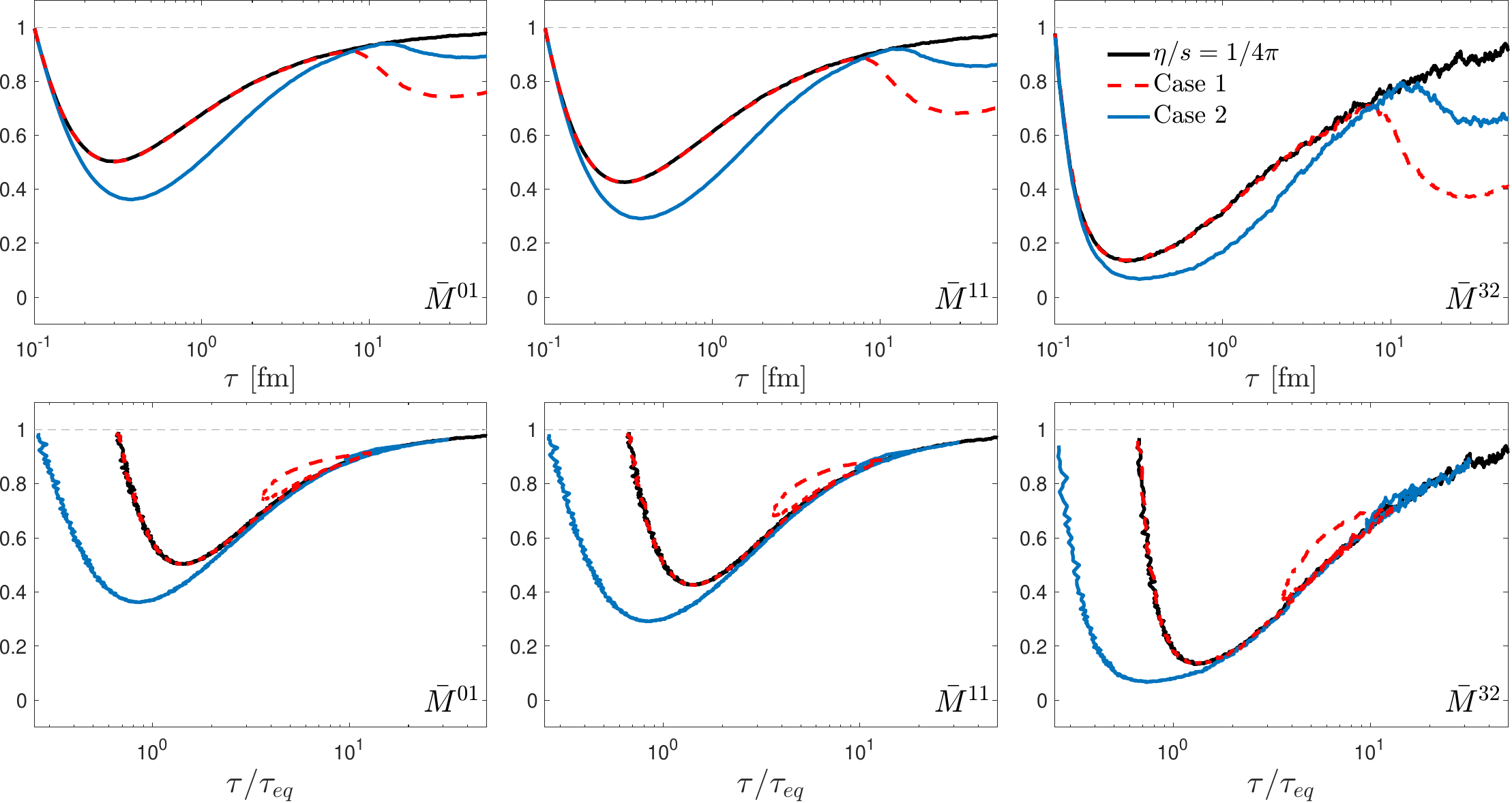}
    \caption{Upper panels: normalized moments as function of $\tau$ at midrapidity and for three different $\eta/s(T)$; colors are the same of Fig. \ref{fig:etas_temp}. Lower panels: normalized moments as function of the scaled time $\tau/\tau_{eq}$. Initial conditions are $T_0 = 0.3$ GeV, $\xi_0=0$. }
    \label{fig:moments_etas_temp}
\end{figure*}
In this section, we extend the previous analysis performed at fixed $\eta/s$, by studying the role of its temperature-dependence. To our knowledge, this has not been studied neither in RTA not in viscous hydrodynamics, despite its relevance for the physical case of Hot QCD matter.
To this end we have considered various parametrization of $\eta/s (T)$, as shown in Fig. \ref{fig:etas_temp} in comparison to the current estimates from lattice QCD \cite{Meyer:2007ic, Borsanyi:2018srz} calculations, chiral perturbation theory ($\chi$PT) \cite{Chen:2007xe}, hadron-resonance gas models and a general Bayesian estimation \cite{Bernhard:2019bmu}. We have considered an extreme case (Case 1) where the specific viscosity changes abruptly at the critical temperature, increasing indefinitely towards smaller temperatures, and a more realistic case (Case 2) with a minimum close to the critical temperature as suggested by combining information from lattice QCD and estimations for the hadronic phase.

In the upper panels of Fig. \ref{fig:moments_etas_temp} we show the time dependence of the moments of the distribution function for the different parametrization of $\eta/s(T)$.
While for fixed specific viscosity the moments increase smoothly with increasing time after the minimum, as seen in detail in the previous sections, this behavior changes when the $\eta/s$ increases at lower temperature. Indeed, at $\tau \approx 2-3$ fm the moments start to have a non-monotonic trend, especially for the Case 1 of $\eta/s(T)$.
This can be understood considering that in our kinetic approach, a change in $\eta/s$ corresponds to a change in the scattering cross section; since collisions drive the system towards the isotropization, a lowering in the scattering cross-section allows for a departure from the previously seen behavior, that is a monotonic increase towards the equilibrium values after the initial free streaming dominance.

In the lower panels of Fig. \ref{fig:moments_etas_temp} the moments are presented as a function of scaled time $\tau/\tau_{eq}$.
In this case we recover the universal scaling at large $\tau/\tau_{eq}$, so that the curves for temperature-dependent $\eta/s$ lie on top of those with fixed specific viscosity.
However, at intermediate values of the scaled time, we notice an interesting departure from the universal behavior. At $\tau/\tau_{eq}\sim 2$-$3$ a small loop is present in the moments calculated with $\eta/s$ that increases at lower temperatures right below the critical value. This region in scaled time corresponds to the temperature region in which $\eta/s$ increases so that $\tau/\tau_{eq}$ is no more a monotonic function of the proper time. {Since, as we know from RTA, $\tau_{eq}\propto (\eta/s) /T$, if $\eta/s$ is an increasing function of the proper time while $T$ decreases, there could be some intervals in which $\tau/\tau_{eq} (\tau)$ is no more monotonic, but instead shows a maximum and a minimum. This can be read out from Fig. \ref{fig:tau_eq_RBT} noticing that the difference between $\tau_{eq}$ and $\tau$ increases and decreases again with time. Specifically for the cases shown in Figure \ref{fig:moments_etas_temp}, one can verify by looking at Fig. \ref{fig:tau_eq_RBT} that the non-monotonic behavior, i.e. the loop in the plots, begin exactly when $\eta/s (T)$ starts increasing. This of course depends on the specific dependence of $T(\tau)$ and $\eta/s (T)$: the steeper the $\eta/s (T)$ curve, the more pronounced the loop. Physically, what happens is that an increasing of the specific shear viscosity make the system recede again from equilibrium, which reflects in a temporary decreasing of the normalized moments and results in a minimum.}
These findings support the idea that an increase of the specific viscosity in the low-temperature region, as expected from lattice QCD and phenomenological calculations, determines a partial breaking of the universal attractor behavior.
It would be interesting to compare these results with hydrodynamic calculations, to evaluate whether even in such a region of $\eta/s(T)$ the behavior is quantitatively similar. From the upper panels of Fig. \ref{fig:moments_etas_temp} we see that an increasing $\eta/s(T)$ for $T<T_c$ drives the system out of equilibrium at $\tau\sim 10$ fm, which however for the realistic matter created at LHC is a time where large part of the freeze-out hypersurface lies. Therefore only a realistic 3+1D simulation can evaluate if such a second receding from equilibrium is expected to occur. For the simpler case of 1+1D expansion one can see a deviation from equilibrium of about 10-15\% for Case 1 and 30\% for Case 2.

\section{Attractors with breaking of boost invariance}
\label{sec:no_boostinv}

The evolution equations of hydrodynamic and RTA used both in our work (see Section \ref{sec:models}) and in previous studies on attractors are boost-invariant.
Our full Boltzmann approach is not inherently boost-invariant, hence we have the advantage to easily explore the impact of the breaking of boost invariance. First of all, it is not granted that a boost-invariant initial condition leads through the dynamical evolution to boost-invariant final phase space distribution functions and thermodynamical quantities, such as temperature ($T$), density ($n$) and energy density ($\varepsilon$).

 \begin{figure}[t!]
     \centering
     \includegraphics[width=\columnwidth]{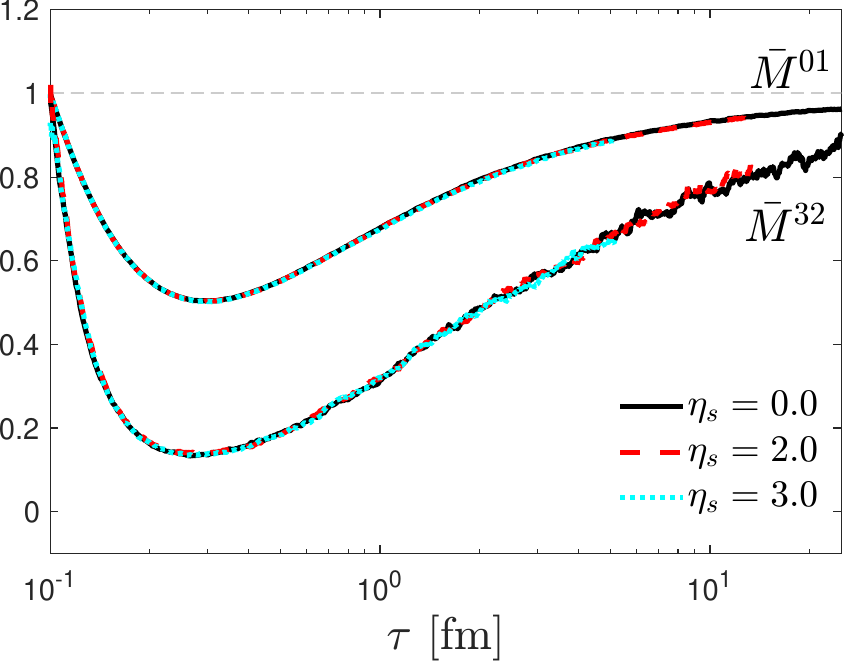}
     \caption{Normalized moments at different rapidities in the case of boost-invariant conditions. Different line styles correspond to: $\eta_s = 0.0$ (black solid), 2.0 (red dashed), 3.0 (cyan dotted). Normalized moments computed at different rapidities perfectly overlap.}
     \label{fig:checking_boost}
 \end{figure}
In this section, we present the results on the attractor behavior obtained with our full Boltzmann approach, first in an  `artificially' boost-invariant case and then in the case where the boost invariance is broken during the dynamics. This is the first time that the impact of the boost invariance breaking on the universal scaling with $\tau_{eq}$ is investigated.

 \begin{figure*}[th!]
     \centering    \includegraphics[width=.95\textwidth]{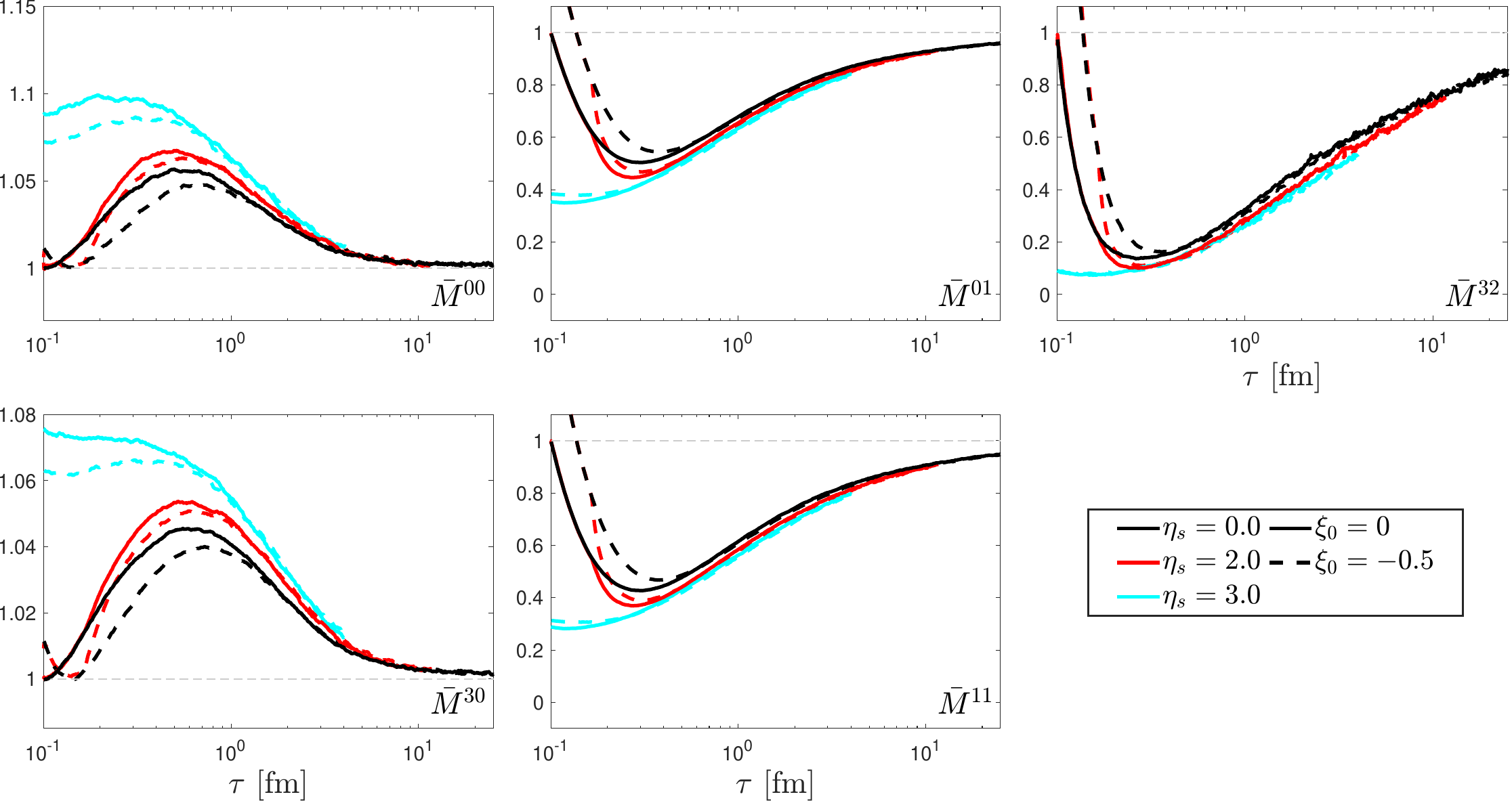}
      \caption{Normalized moments as a function of $\tau$ within the RBT approach. Different styles correspond to two different values of initial anisotropy $\xi_0 = [-0.5,\,0]$, highlighting the forward attractor behavior. The three colors correspond to different rapidities: $\eta_s = 0.0$ (black), 2.0 (red), 3.0 (cyan). We fix $T_0= 0.5$ GeV and $\tau_0 = 0.1$ fm.}
     \label{fig:forwardrap_non-Bj_forward}
\end{figure*}
In our approach the boost-invariant dynamics is achieved simulating a system with a large extension in the longitudinal direction with respect to the region which we are interested in. In this way, the information from the surface does not propagate to the considered region. 

In Fig. \ref{fig:checking_boost} we show two moments of the distribution function versus proper time for different values of the space-time rapidity.
The initial longitudinal extension of the system is $|\eta|<8.5$ and the observed rapidities are $\eta=[0,\,2,\,3]$. 
We see that all curves lie one on top of each other, indicating that the attractor behavior seen at midrapidity is maintained when looking to the system at non-central rapidity. We have checked that the same scaling is observed for the other momentum moments not shown in the figure and for different values of initial anisotropies and viscosity.
As mentioned before, this case is with excellent approximation boost-invariant. Indeed, we have checked that $T$, $n$ and $\varepsilon$ do not differ for the considered values of $\eta$ during the whole time evolution. By computing the fluid four-velocity $u^\mu$ as the eigenvector of the energy-momentum tensor, we find that, for each cell, it fulfils the relation $u^\mu =(\cosh\eta ,0,0,\sinh\eta)$, 
which, according to Eq. (\ref{eq:umu_bjorken}), is exactly the boost-invariant condition. It means of course that the Lorentz factor $\gamma=\cosh\eta$ and $\gamma \beta_z = \sinh(\eta)$.

\begin{figure*}[th!]
    \centering  \includegraphics[width=.95\textwidth]
{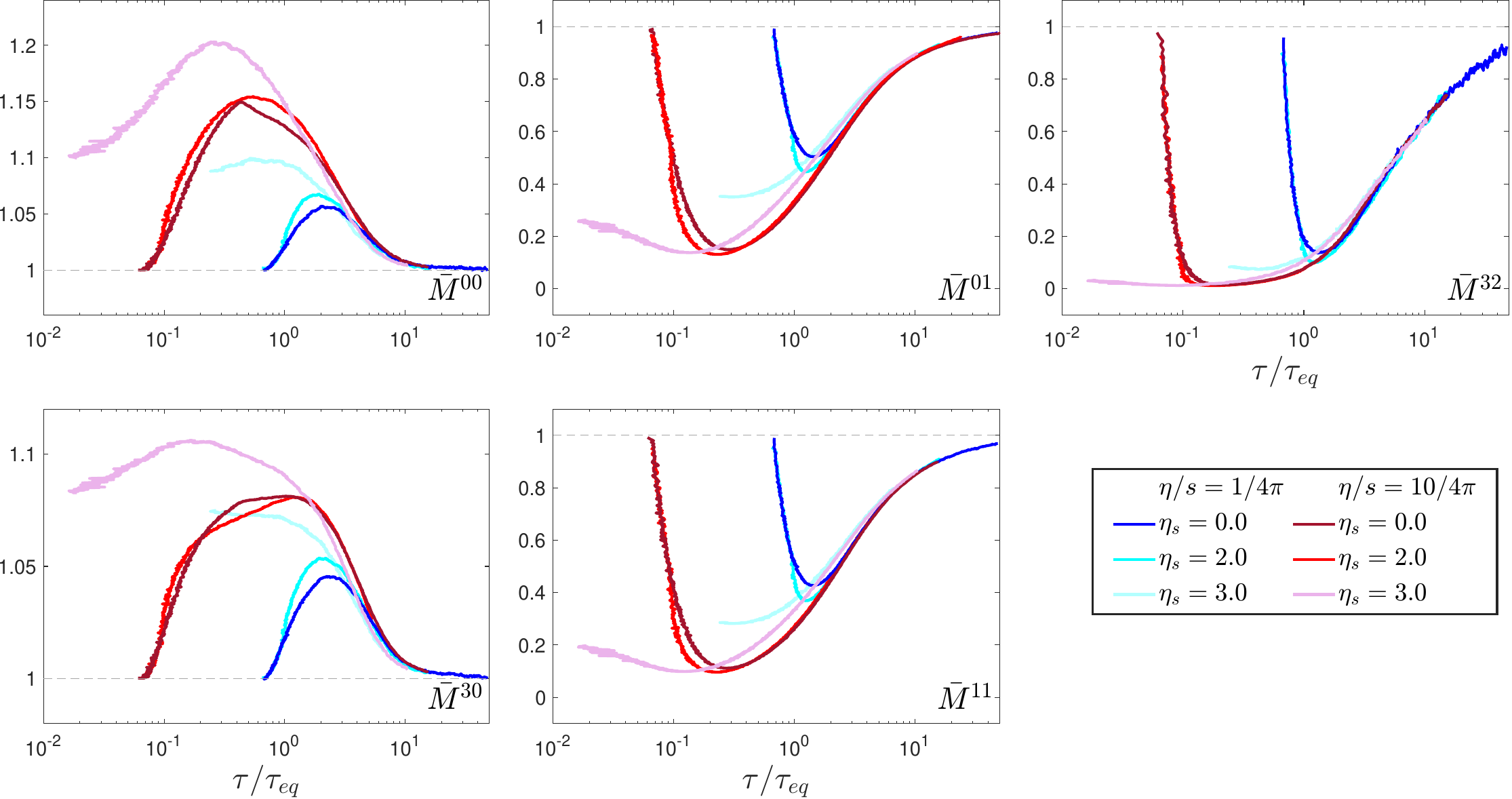}
     \caption{Normalized moments as a function of $\tau/\tau_{eq}$ within the RBT approach. Different color scales correspond to two different values of $\eta/s$ highlighting the pull-back attractor behavior. The three shades correspond to different rapidities: $\eta_s = 0.0,\, 2.0,\, 3.0$. We fix $T_0= 0.5$ GeV and $\tau_0 = 0.1$ fm.}
     \label{fig:forwardrap_non-Bj_pullback}
 \end{figure*}

We present now the results obtained with simulations in which the initial longitudinal extension of the system is limited ($|\eta|<2.5$), so that the information from the external surface of the system reaches the $\eta$ region under consideration, leading in practice to a breaking of the boost-invariance during the dynamical evolution.
This is visible by looking at $T$, $n$ and $\varepsilon$ that show a different time decay as a function of the space-time rapidity.
Moreover, the boost-invariance breaking implies that $\gamma\neq\cosh\eta$. This proves that the fluid is no more in Bjorken flow, and therefore $u^\mu$ and $z^\mu$ are no more related with $\cosh\eta$ and $\sinh\eta$. This means that we cannot use anymore the expressions in Eq. \ref{eq:strick_moments}, but compute directly $p^w=p\cdot z$ and $p^\tau = p\cdot u$ and use these quantities in the calculation of the momentum moments in Eq. \eqref{eq:momentum_moments}.
In Fig. \ref{fig:forwardrap_non-Bj_forward} we show some of the moments at different values of the space-time rapidity as a function of the proper time $\tau$.
The various curves correspond to $\eta/s=1/(4 \pi)$ and different values of the initial anisotropy $\xi_0$; we clearly see that the so called forward attractor behavior emerges regardless of the rapidity considered.
In Fig. \ref{fig:forwardrap_non-Bj_pullback} the simulations have been performed with $\xi_0=0$ and for different values of the specific viscosity; even in this case we see that the universal scaling with respect to the scaled time $\tau/\tau_{eq}$ is reached and the pull-back attractor does not depend on the space-time rapidity.

It is remarkable that the loss of information driving the system towards the attractor is effective also for $\eta=3.0$, where at initial time there are no particles due to our initialization limited to $|\eta|<2.5$. Once such space-time region is populated due to the momentum transferred between particles by means of collisions, the corresponding distribution function moment approaches the attractor.

\section{Conclusion}
\label{sec:concl}

In this paper we have studied the dynamical attractors associated with the full Relativistic Boltzmann Transport (RBT) approach studying several order moments of the one-particle distribution function and the full phase-space of the distribution function itself. We have investigated two different cases: in the first one we maintain the longitudinal boost invariance, as similarly done currently in the literature in $(0+1)$D hydrodynamic models and the RTA Boltzmann equation; in the second case the boost-invariance is broken dynamically during the evolution. 
For the first case we have demonstrated that the resulting solutions of RBT converge to a unique attractor even though the system is out of chemical equilibrium with $\Gamma \neq 1$.
Similarly to what has been pointed out in previous studies, we clearly observe also in the RBT approach that the dynamical evolution is initially governed by free-streaming, which leads to a deviation of the distribution function moments from their equilibrium value; subsequently, the collisional dynamics starts to play a role driving the systems towards the equilibrium limit. 
{{We identify that in our RBT framework the timescale in which the system approaches the equilibrium and reaches the universal attractor starting from different initial conditions is related to the collision rate.}}
This is manifest in the observation of both pull-back and forward attractors of the momentum moments when the proper time is scaled according to this quantity.
We also compared the resulting attractor moments with calculations from different hydrodynamic frameworks, such as aHydro and DNMR, as well as from kinetic theory in RTA. We show that the same attractor solution for lower momentum moments between models is obtained and a difference emerges for $m,n \ge 2$. More specifically, RBT exhibits an even faster equilibration of the higher moments, linked to higher $p_T$ region of the distribution function. At times comparable to the QGP lifetime in $AA$ collisions at LHC one gets a nearly full thermalization for $\overline{M}^{mn}$ at least up to $m,n=3$, while RTA, aHydro, DNMR have value about $15-20\%$ smaller.
{However, we observe that, by construction, RTA and aHydro show better agreement with the RBT approach for a wider range of normalized moments.}\\
In our RBT approach it is also possible to extend the analysis by considering a more realistic scenario with a temperature-dependent $\eta/s(T)$ that is relevant to the realistic case of Hot QCD matter, where $\eta/s(T)$ is expected to smoothly decrease with decreasing $T$ down to about $T_c$ and then to have a rapid increase at lower temperature corresponding to the hadronic phase. To our knowledge this is the first study exploring such a case.
We have considered different parametrizations including also the extreme case of a sudden freeze-out. In all the cases studied we observe that the universal scaling is obtained at large scaled times, but a partial breaking emerges at intermediate time corresponding to lower temperatures in which the specific viscosity increases.
However, such a breaking is restricted to a limited region of temperature and here the moments $\overline{M}^{mn}$  as a function of
$\tau/\tau_{eq}$ loop around the attractor but then lie again along it (see lower panels of Fig. \ref{fig:moments_etas_temp}).
On the other hand it is interesting to notice that due to the strong increase of $\eta/s$ below $T_c$ this is achieved thanks to a larger increase of $\tau_{eq}$.
In fact the rise of $\eta/s(T)$ in the hadronic phase implies a very fast
increase from $\tau_{eq} \simeq  0.5$ fm at $\tau < 10$ fm up to an order of magnitude, $\tau_{eq} \simeq  5$ fm,  at $\tau \simeq 15$ fm (Fig. \ref{fig:tau_eq_RBT}).
Therefore, as a function of temperature one can notice that $\eta/s(T)$ induces a receding from full thermalization at $\tau \sim \, 10-20\rm \,fm$, which are comparable to those where the freeze-out hypersurface in $AA$ collisions lies. In fact in the upper panels of Fig. \ref{fig:etas_temp} one can see that for times typical of QGP lifetime the system can remain significantly out of equilibrium, especially in case of a quite rapid increase of $\eta/s(T)$ for $T<T_c$. 
This is particularly interesting because the kinetic freeze-out dynamics is by 
definition opposite to hydrodynamics assumptions and one would associate it to
a rising $\eta/s$ or similarly to a vanishing collision rate.
However, the real impact of such a dynamics induced by raising $\eta/s(T)$ has to be evaluated in 3+1D under realistic conditions and a quantitative assessment has to be elaborated, which can be achieved in a natural extension of the present RBT approach.
Furthermore this could have an even stronger impact on the understanding of small systems like $pp$ and $pA$.

Finally, we have investigated for the first time the behavior of the normalized moments at larger space-time rapidity. We find that the same universal scaling observed at midrapidity is maintained at larger rapidity when the system evolves through the Bjorken flow. However, we find for the first time that a universal behavior is observed in the momentum moments of the phase-space distribution function also when the longitudinal boost invariance is broken.\\
The present work has been mainly devoted to the presentation of the RBT approach and the comparison to existing approaches like aHydro, DNMR, RTA
largely employed to study the 0+1 D dynamics for conformal systems under the Bjorken scaling condition for a fluid at constant $\eta/s$. However, we have shown that RBT is quite flexible and can be easily extended to investigate the fluid dynamics in the case of $T$-dependent $\eta/s$ and of broken Bjorken symmetry, as discussed above.  The real advantage of RBT will be the extension to 3+1D dynamics for both conformal and non conformal dynamics, which will allow us also to study the dynamics of anisotropic flows $v_n= \langle\cos(n\,\varphi_p)\rangle$, which is currently under development.

\subsection*{Acknowledgments}

We thank M. Strickland for enlightening discussions. V.G. acknowledges the funding from UniCT under ‘Linea di intervento 2’ (HQCDyn Grant).
L.O. acknowledges the Next Generation action of the European Commission and the MUR funding (PNRR Missione 4) under the HEFESTUS project.
This work is (partially) supported by ICSC – Centro Nazionale di Ricerca in High Performance Computing, Big Data and Quantum Computing, funded by European Union – NextGenerationEU.


\begin{thebibliography}{79}
\expandafter\ifx\csname natexlab\endcsname\relax\def\natexlab#1{#1}\fi
\expandafter\ifx\csname bibnamefont\endcsname\relax
  \def\bibnamefont#1{#1}\fi
\expandafter\ifx\csname bibfnamefont\endcsname\relax
  \def\bibfnamefont#1{#1}\fi
\expandafter\ifx\csname citenamefont\endcsname\relax
  \def\citenamefont#1{#1}\fi
\expandafter\ifx\csname url\endcsname\relax
  \def\url#1{\texttt{#1}}\fi
\expandafter\ifx\csname urlprefix\endcsname\relax\def\urlprefix{URL }\fi
\providecommand{\bibinfo}[2]{#2}
\providecommand{\eprint}[2][]{\url{#2}}

\bibitem[{\citenamefont{Romatschke and Romatschke}(2019)}]{Romatschke:2017ejr}
\bibinfo{author}{\bibfnamefont{P.}~\bibnamefont{Romatschke}} \bibnamefont{and}
  \bibinfo{author}{\bibfnamefont{U.}~\bibnamefont{Romatschke}},
  \emph{\bibinfo{title}{{Relativistic Fluid Dynamics In and Out of
  Equilibrium}}}, Cambridge Monographs on Mathematical Physics
  (\bibinfo{publisher}{Cambridge University Press}, \bibinfo{year}{2019}), ISBN
  \bibinfo{isbn}{978-1-108-48368-1, 978-1-108-75002-8}, \eprint{1712.05815}.

\bibitem[{\citenamefont{Florkowski et~al.}(2018)\citenamefont{Florkowski,
  Heller, and Spalinski}}]{Florkowski:2017olj}
\bibinfo{author}{\bibfnamefont{W.}~\bibnamefont{Florkowski}},
  \bibinfo{author}{\bibfnamefont{M.~P.} \bibnamefont{Heller}},
  \bibnamefont{and}
  \bibinfo{author}{\bibfnamefont{M.}~\bibnamefont{Spalinski}},
  \bibinfo{journal}{Rept. Prog. Phys.} \textbf{\bibinfo{volume}{81}},
  \bibinfo{pages}{046001} (\bibinfo{year}{2018}), \eprint{1707.02282}.

\bibitem[{\citenamefont{Heinz and Snellings}(2013)}]{Heinz:2013th}
\bibinfo{author}{\bibfnamefont{U.}~\bibnamefont{Heinz}} \bibnamefont{and}
  \bibinfo{author}{\bibfnamefont{R.}~\bibnamefont{Snellings}},
  \bibinfo{journal}{Ann. Rev. Nucl. Part. Sci.} \textbf{\bibinfo{volume}{63}},
  \bibinfo{pages}{123} (\bibinfo{year}{2013}), \eprint{1301.2826}.

\bibitem[{\citenamefont{Denicol et~al.}(2012)\citenamefont{Denicol, Niemi,
  Molnar, and Rischke}}]{Denicol:2012cn}
\bibinfo{author}{\bibfnamefont{G.~S.} \bibnamefont{Denicol}},
  \bibinfo{author}{\bibfnamefont{H.}~\bibnamefont{Niemi}},
  \bibinfo{author}{\bibfnamefont{E.}~\bibnamefont{Molnar}}, \bibnamefont{and}
  \bibinfo{author}{\bibfnamefont{D.~H.} \bibnamefont{Rischke}},
  \bibinfo{journal}{Phys. Rev. D} \textbf{\bibinfo{volume}{85}},
  \bibinfo{pages}{114047} (\bibinfo{year}{2012}), \bibinfo{note}{[Erratum:
  Phys.Rev.D 91, 039902 (2015)]}, \eprint{1202.4551}.

\bibitem[{\citenamefont{Ruggieri et~al.}(2013)\citenamefont{Ruggieri, Scardina,
  Plumari, and Greco}}]{Ruggieri:2013bda}
\bibinfo{author}{\bibfnamefont{M.}~\bibnamefont{Ruggieri}},
  \bibinfo{author}{\bibfnamefont{F.}~\bibnamefont{Scardina}},
  \bibinfo{author}{\bibfnamefont{S.}~\bibnamefont{Plumari}}, \bibnamefont{and}
  \bibinfo{author}{\bibfnamefont{V.}~\bibnamefont{Greco}},
  \bibinfo{journal}{Phys. Lett. B} \textbf{\bibinfo{volume}{727}},
  \bibinfo{pages}{177} (\bibinfo{year}{2013}), \eprint{1303.3178}.

\bibitem[{\citenamefont{Ruggieri et~al.}(2014)\citenamefont{Ruggieri, Scardina,
  Plumari, and Greco}}]{Ruggieri:2013ova}
\bibinfo{author}{\bibfnamefont{M.}~\bibnamefont{Ruggieri}},
  \bibinfo{author}{\bibfnamefont{F.}~\bibnamefont{Scardina}},
  \bibinfo{author}{\bibfnamefont{S.}~\bibnamefont{Plumari}}, \bibnamefont{and}
  \bibinfo{author}{\bibfnamefont{V.}~\bibnamefont{Greco}},
  \bibinfo{journal}{Phys. Rev. C} \textbf{\bibinfo{volume}{89}},
  \bibinfo{pages}{054914} (\bibinfo{year}{2014}), \eprint{1312.6060}.

\bibitem[{\citenamefont{Ruggieri et~al.}(2015)\citenamefont{Ruggieri, Puglisi,
  Oliva, Plumari, Scardina, and Greco}}]{Ruggieri:2015yea}
\bibinfo{author}{\bibfnamefont{M.}~\bibnamefont{Ruggieri}},
  \bibinfo{author}{\bibfnamefont{A.}~\bibnamefont{Puglisi}},
  \bibinfo{author}{\bibfnamefont{L.}~\bibnamefont{Oliva}},
  \bibinfo{author}{\bibfnamefont{S.}~\bibnamefont{Plumari}},
  \bibinfo{author}{\bibfnamefont{F.}~\bibnamefont{Scardina}}, \bibnamefont{and}
  \bibinfo{author}{\bibfnamefont{V.}~\bibnamefont{Greco}},
  \bibinfo{journal}{Phys. Rev. C} \textbf{\bibinfo{volume}{92}},
  \bibinfo{pages}{064904} (\bibinfo{year}{2015}), \eprint{1505.08081}.

\bibitem[{\citenamefont{Plumari
  et~al.}(2015{\natexlab{a}})\citenamefont{Plumari, Guardo, Greco, and
  Ollitrault}}]{Plumari:2015sia}
\bibinfo{author}{\bibfnamefont{S.}~\bibnamefont{Plumari}},
  \bibinfo{author}{\bibfnamefont{G.~L.} \bibnamefont{Guardo}},
  \bibinfo{author}{\bibfnamefont{V.}~\bibnamefont{Greco}}, \bibnamefont{and}
  \bibinfo{author}{\bibfnamefont{J.-Y.} \bibnamefont{Ollitrault}},
  \bibinfo{journal}{Nucl. Phys. A} \textbf{\bibinfo{volume}{941}},
  \bibinfo{pages}{87} (\bibinfo{year}{2015}{\natexlab{a}}),
  \eprint{1502.04066}.

\bibitem[{\citenamefont{Plumari}(2019)}]{Plumari:2019gwq}
\bibinfo{author}{\bibfnamefont{S.}~\bibnamefont{Plumari}},
  \bibinfo{journal}{Eur. Phys. J. C} \textbf{\bibinfo{volume}{79}},
  \bibinfo{pages}{2} (\bibinfo{year}{2019}).

\bibitem[{\citenamefont{Xu and Greiner}(2005)}]{Xu:2004mz}
\bibinfo{author}{\bibfnamefont{Z.}~\bibnamefont{Xu}} \bibnamefont{and}
  \bibinfo{author}{\bibfnamefont{C.}~\bibnamefont{Greiner}},
  \bibinfo{journal}{Phys. Rev. C} \textbf{\bibinfo{volume}{71}},
  \bibinfo{pages}{064901} (\bibinfo{year}{2005}), \eprint{hep-ph/0406278}.

\bibitem[{\citenamefont{Uphoff et~al.}(2015)\citenamefont{Uphoff, Senzel,
  Fochler, Wesp, Xu, and Greiner}}]{Uphoff:2014cba}
\bibinfo{author}{\bibfnamefont{J.}~\bibnamefont{Uphoff}},
  \bibinfo{author}{\bibfnamefont{F.}~\bibnamefont{Senzel}},
  \bibinfo{author}{\bibfnamefont{O.}~\bibnamefont{Fochler}},
  \bibinfo{author}{\bibfnamefont{C.}~\bibnamefont{Wesp}},
  \bibinfo{author}{\bibfnamefont{Z.}~\bibnamefont{Xu}}, \bibnamefont{and}
  \bibinfo{author}{\bibfnamefont{C.}~\bibnamefont{Greiner}},
  \bibinfo{journal}{Phys. Rev. Lett.} \textbf{\bibinfo{volume}{114}},
  \bibinfo{pages}{112301} (\bibinfo{year}{2015}), \eprint{1401.1364}.

\bibitem[{\citenamefont{Cassing and Bratkovskaya}(2009)}]{Cassing:2009vt}
\bibinfo{author}{\bibfnamefont{W.}~\bibnamefont{Cassing}} \bibnamefont{and}
  \bibinfo{author}{\bibfnamefont{E.~L.} \bibnamefont{Bratkovskaya}},
  \bibinfo{journal}{Nucl. Phys. A} \textbf{\bibinfo{volume}{831}},
  \bibinfo{pages}{215} (\bibinfo{year}{2009}), \eprint{0907.5331}.

\bibitem[{\citenamefont{Bratkovskaya et~al.}(2011)\citenamefont{Bratkovskaya,
  Cassing, Konchakovski, and Linnyk}}]{Bratkovskaya:2011wp}
\bibinfo{author}{\bibfnamefont{E.~L.} \bibnamefont{Bratkovskaya}},
  \bibinfo{author}{\bibfnamefont{W.}~\bibnamefont{Cassing}},
  \bibinfo{author}{\bibfnamefont{V.~P.} \bibnamefont{Konchakovski}},
  \bibnamefont{and} \bibinfo{author}{\bibfnamefont{O.}~\bibnamefont{Linnyk}},
  \bibinfo{journal}{Nucl. Phys. A} \textbf{\bibinfo{volume}{856}},
  \bibinfo{pages}{162} (\bibinfo{year}{2011}), \eprint{1101.5793}.

\bibitem[{\citenamefont{Soloveva et~al.}(2022)\citenamefont{Soloveva, Aichelin,
  and Bratkovskaya}}]{Soloveva:2021quj}
\bibinfo{author}{\bibfnamefont{O.}~\bibnamefont{Soloveva}},
  \bibinfo{author}{\bibfnamefont{J.}~\bibnamefont{Aichelin}}, \bibnamefont{and}
  \bibinfo{author}{\bibfnamefont{E.}~\bibnamefont{Bratkovskaya}},
  \bibinfo{journal}{Phys. Rev. D} \textbf{\bibinfo{volume}{105}},
  \bibinfo{pages}{054011} (\bibinfo{year}{2022}), \eprint{2108.08561}.

\bibitem[{\citenamefont{Kurkela et~al.}(2019)\citenamefont{Kurkela,
  Mazeliauskas, Paquet, Schlichting, and Teaney}}]{Kurkela:2018vqr}
\bibinfo{author}{\bibfnamefont{A.}~\bibnamefont{Kurkela}},
  \bibinfo{author}{\bibfnamefont{A.}~\bibnamefont{Mazeliauskas}},
  \bibinfo{author}{\bibfnamefont{J.-F.} \bibnamefont{Paquet}},
  \bibinfo{author}{\bibfnamefont{S.}~\bibnamefont{Schlichting}},
  \bibnamefont{and} \bibinfo{author}{\bibfnamefont{D.}~\bibnamefont{Teaney}},
  \bibinfo{journal}{Phys. Rev. C} \textbf{\bibinfo{volume}{99}},
  \bibinfo{pages}{034910} (\bibinfo{year}{2019}), \eprint{1805.00961}.

\bibitem[{\citenamefont{Oliva et~al.}(2022)\citenamefont{Oliva, Fan, Moreau,
  Bass, and Bratkovskaya}}]{Oliva:2022rsv}
\bibinfo{author}{\bibfnamefont{L.}~\bibnamefont{Oliva}},
  \bibinfo{author}{\bibfnamefont{W.}~\bibnamefont{Fan}},
  \bibinfo{author}{\bibfnamefont{P.}~\bibnamefont{Moreau}},
  \bibinfo{author}{\bibfnamefont{S.~A.} \bibnamefont{Bass}}, \bibnamefont{and}
  \bibinfo{author}{\bibfnamefont{E.}~\bibnamefont{Bratkovskaya}},
  \bibinfo{journal}{Phys. Rev. C} \textbf{\bibinfo{volume}{106}},
  \bibinfo{pages}{044910} (\bibinfo{year}{2022}), \eprint{2204.04194}.

\bibitem[{\citenamefont{Yan and Ollitrault}(2014)}]{Yan:2013laa}
\bibinfo{author}{\bibfnamefont{L.}~\bibnamefont{Yan}} \bibnamefont{and}
  \bibinfo{author}{\bibfnamefont{J.-Y.} \bibnamefont{Ollitrault}},
  \bibinfo{journal}{Phys. Rev. Lett.} \textbf{\bibinfo{volume}{112}},
  \bibinfo{pages}{082301} (\bibinfo{year}{2014}), \eprint{1312.6555}.

\bibitem[{\citenamefont{Romatschke}(2015)}]{Romatschke:2015gxa}
\bibinfo{author}{\bibfnamefont{P.}~\bibnamefont{Romatschke}},
  \bibinfo{journal}{Eur. Phys. J. C} \textbf{\bibinfo{volume}{75}},
  \bibinfo{pages}{305} (\bibinfo{year}{2015}), \eprint{1502.04745}.

\bibitem[{\citenamefont{Shen et~al.}(2017)\citenamefont{Shen, Paquet, Denicol,
  Jeon, and Gale}}]{Shen:2016zpp}
\bibinfo{author}{\bibfnamefont{C.}~\bibnamefont{Shen}},
  \bibinfo{author}{\bibfnamefont{J.-F.} \bibnamefont{Paquet}},
  \bibinfo{author}{\bibfnamefont{G.~S.} \bibnamefont{Denicol}},
  \bibinfo{author}{\bibfnamefont{S.}~\bibnamefont{Jeon}}, \bibnamefont{and}
  \bibinfo{author}{\bibfnamefont{C.}~\bibnamefont{Gale}},
  \bibinfo{journal}{Phys. Rev. C} \textbf{\bibinfo{volume}{95}},
  \bibinfo{pages}{014906} (\bibinfo{year}{2017}), \eprint{1609.02590}.

\bibitem[{\citenamefont{M\"antysaari et~al.}(2017)\citenamefont{M\"antysaari,
  Schenke, Shen, and Tribedy}}]{Mantysaari:2017cni}
\bibinfo{author}{\bibfnamefont{H.}~\bibnamefont{M\"antysaari}},
  \bibinfo{author}{\bibfnamefont{B.}~\bibnamefont{Schenke}},
  \bibinfo{author}{\bibfnamefont{C.}~\bibnamefont{Shen}}, \bibnamefont{and}
  \bibinfo{author}{\bibfnamefont{P.}~\bibnamefont{Tribedy}},
  \bibinfo{journal}{Phys. Lett. B} \textbf{\bibinfo{volume}{772}},
  \bibinfo{pages}{681} (\bibinfo{year}{2017}), \eprint{1705.03177}.

\bibitem[{\citenamefont{Zhao et~al.}(2023)\citenamefont{Zhao, Ryu, Shen, and
  Schenke}}]{Zhao:2022ugy}
\bibinfo{author}{\bibfnamefont{W.}~\bibnamefont{Zhao}},
  \bibinfo{author}{\bibfnamefont{S.}~\bibnamefont{Ryu}},
  \bibinfo{author}{\bibfnamefont{C.}~\bibnamefont{Shen}}, \bibnamefont{and}
  \bibinfo{author}{\bibfnamefont{B.}~\bibnamefont{Schenke}},
  \bibinfo{journal}{Phys. Rev. C} \textbf{\bibinfo{volume}{107}},
  \bibinfo{pages}{014904} (\bibinfo{year}{2023}), \eprint{2211.16376}.

\bibitem[{\citenamefont{Heller and Spalinski}(2015)}]{Heller:2015dha}
\bibinfo{author}{\bibfnamefont{M.~P.} \bibnamefont{Heller}} \bibnamefont{and}
  \bibinfo{author}{\bibfnamefont{M.}~\bibnamefont{Spalinski}},
  \bibinfo{journal}{Phys. Rev. Lett.} \textbf{\bibinfo{volume}{115}},
  \bibinfo{pages}{072501} (\bibinfo{year}{2015}), \eprint{1503.07514}.

\bibitem[{\citenamefont{Strickland et~al.}(2018)\citenamefont{Strickland,
  Noronha, and Denicol}}]{Strickland:2017kux}
\bibinfo{author}{\bibfnamefont{M.}~\bibnamefont{Strickland}},
  \bibinfo{author}{\bibfnamefont{J.}~\bibnamefont{Noronha}}, \bibnamefont{and}
  \bibinfo{author}{\bibfnamefont{G.}~\bibnamefont{Denicol}},
  \bibinfo{journal}{Phys. Rev. D} \textbf{\bibinfo{volume}{97}},
  \bibinfo{pages}{036020} (\bibinfo{year}{2018}), \eprint{1709.06644}.

\bibitem[{\citenamefont{Chattopadhyay and Heinz}(2020)}]{Chattopadhyay:2019jqj}
\bibinfo{author}{\bibfnamefont{C.}~\bibnamefont{Chattopadhyay}}
  \bibnamefont{and} \bibinfo{author}{\bibfnamefont{U.~W.} \bibnamefont{Heinz}},
  \bibinfo{journal}{Phys. Lett. B} \textbf{\bibinfo{volume}{801}},
  \bibinfo{pages}{135158} (\bibinfo{year}{2020}), \eprint{1911.07765}.

\bibitem[{\citenamefont{Jaiswal et~al.}(2019)\citenamefont{Jaiswal,
  Chattopadhyay, Jaiswal, Pal, and Heinz}}]{Jaiswal:2019cju}
\bibinfo{author}{\bibfnamefont{S.}~\bibnamefont{Jaiswal}},
  \bibinfo{author}{\bibfnamefont{C.}~\bibnamefont{Chattopadhyay}},
  \bibinfo{author}{\bibfnamefont{A.}~\bibnamefont{Jaiswal}},
  \bibinfo{author}{\bibfnamefont{S.}~\bibnamefont{Pal}}, \bibnamefont{and}
  \bibinfo{author}{\bibfnamefont{U.}~\bibnamefont{Heinz}},
  \bibinfo{journal}{Phys. Rev. C} \textbf{\bibinfo{volume}{100}},
  \bibinfo{pages}{034901} (\bibinfo{year}{2019}), \eprint{1907.07965}.

\bibitem[{\citenamefont{Blaizot and Yan}(2020)}]{Blaizot:2019scw}
\bibinfo{author}{\bibfnamefont{J.-P.} \bibnamefont{Blaizot}} \bibnamefont{and}
  \bibinfo{author}{\bibfnamefont{L.}~\bibnamefont{Yan}},
  \bibinfo{journal}{Annals Phys.} \textbf{\bibinfo{volume}{412}},
  \bibinfo{pages}{167993} (\bibinfo{year}{2020}), \eprint{1904.08677}.

\bibitem[{\citenamefont{Alalawi and Strickland}(2020)}]{Alalawi:2020zbx}
\bibinfo{author}{\bibfnamefont{H.}~\bibnamefont{Alalawi}} \bibnamefont{and}
  \bibinfo{author}{\bibfnamefont{M.}~\bibnamefont{Strickland}},
  \bibinfo{journal}{Phys. Rev. C} \textbf{\bibinfo{volume}{102}},
  \bibinfo{pages}{064904} (\bibinfo{year}{2020}), \eprint{2006.13834}.

\bibitem[{\citenamefont{Heller et~al.}(2020)\citenamefont{Heller, Jefferson,
  Spali\'nski, and Svensson}}]{Heller:2020anv}
\bibinfo{author}{\bibfnamefont{M.~P.} \bibnamefont{Heller}},
  \bibinfo{author}{\bibfnamefont{R.}~\bibnamefont{Jefferson}},
  \bibinfo{author}{\bibfnamefont{M.}~\bibnamefont{Spali\'nski}},
  \bibnamefont{and} \bibinfo{author}{\bibfnamefont{V.}~\bibnamefont{Svensson}},
  \bibinfo{journal}{Phys. Rev. Lett.} \textbf{\bibinfo{volume}{125}},
  \bibinfo{pages}{132301} (\bibinfo{year}{2020}), \eprint{2003.07368}.

\bibitem[{\citenamefont{Kurkela et~al.}(2020)\citenamefont{Kurkela, van~der
  Schee, Wiedemann, and Wu}}]{Kurkela:2019set}
\bibinfo{author}{\bibfnamefont{A.}~\bibnamefont{Kurkela}},
  \bibinfo{author}{\bibfnamefont{W.}~\bibnamefont{van~der Schee}},
  \bibinfo{author}{\bibfnamefont{U.~A.} \bibnamefont{Wiedemann}},
  \bibnamefont{and} \bibinfo{author}{\bibfnamefont{B.}~\bibnamefont{Wu}},
  \bibinfo{journal}{Phys. Rev. Lett.} \textbf{\bibinfo{volume}{124}},
  \bibinfo{pages}{102301} (\bibinfo{year}{2020}), \eprint{1907.08101}.

\bibitem[{\citenamefont{Almaalol et~al.}(2020)\citenamefont{Almaalol, Kurkela,
  and Strickland}}]{Almaalol:2020rnu}
\bibinfo{author}{\bibfnamefont{D.}~\bibnamefont{Almaalol}},
  \bibinfo{author}{\bibfnamefont{A.}~\bibnamefont{Kurkela}}, \bibnamefont{and}
  \bibinfo{author}{\bibfnamefont{M.}~\bibnamefont{Strickland}},
  \bibinfo{journal}{Phys. Rev. Lett.} \textbf{\bibinfo{volume}{125}},
  \bibinfo{pages}{122302} (\bibinfo{year}{2020}), \eprint{2004.05195}.

\bibitem[{\citenamefont{Behtash et~al.}(2018)\citenamefont{Behtash,
  Cruz-Camacho, and Martinez}}]{Behtash:2017wqg}
\bibinfo{author}{\bibfnamefont{A.}~\bibnamefont{Behtash}},
  \bibinfo{author}{\bibfnamefont{C.~N.} \bibnamefont{Cruz-Camacho}},
  \bibnamefont{and} \bibinfo{author}{\bibfnamefont{M.}~\bibnamefont{Martinez}},
  \bibinfo{journal}{Phys. Rev. D} \textbf{\bibinfo{volume}{97}},
  \bibinfo{pages}{044041} (\bibinfo{year}{2018}), \eprint{1711.01745}.

\bibitem[{\citenamefont{Blaizot and Yan}(2018)}]{Blaizot:2017ucy}
\bibinfo{author}{\bibfnamefont{J.-P.} \bibnamefont{Blaizot}} \bibnamefont{and}
  \bibinfo{author}{\bibfnamefont{L.}~\bibnamefont{Yan}},
  \bibinfo{journal}{Phys. Lett. B} \textbf{\bibinfo{volume}{780}},
  \bibinfo{pages}{283} (\bibinfo{year}{2018}), \eprint{1712.03856}.

\bibitem[{\citenamefont{Strickland}(2018)}]{Strickland:2018ayk}
\bibinfo{author}{\bibfnamefont{M.}~\bibnamefont{Strickland}},
  \bibinfo{journal}{JHEP} \textbf{\bibinfo{volume}{12}}, \bibinfo{pages}{128}
  (\bibinfo{year}{2018}), \eprint{1809.01200}.

\bibitem[{\citenamefont{Heller and Svensson}(2018)}]{Heller:2018qvh}
\bibinfo{author}{\bibfnamefont{M.~P.} \bibnamefont{Heller}} \bibnamefont{and}
  \bibinfo{author}{\bibfnamefont{V.}~\bibnamefont{Svensson}},
  \bibinfo{journal}{Phys. Rev. D} \textbf{\bibinfo{volume}{98}},
  \bibinfo{pages}{054016} (\bibinfo{year}{2018}), \eprint{1802.08225}.

\bibitem[{\citenamefont{Kamata et~al.}(2020)\citenamefont{Kamata, Martinez,
  Plaschke, Ochsenfeld, and Schlichting}}]{Kamata:2020mka}
\bibinfo{author}{\bibfnamefont{S.}~\bibnamefont{Kamata}},
  \bibinfo{author}{\bibfnamefont{M.}~\bibnamefont{Martinez}},
  \bibinfo{author}{\bibfnamefont{P.}~\bibnamefont{Plaschke}},
  \bibinfo{author}{\bibfnamefont{S.}~\bibnamefont{Ochsenfeld}},
  \bibnamefont{and}
  \bibinfo{author}{\bibfnamefont{S.}~\bibnamefont{Schlichting}},
  \bibinfo{journal}{Phys. Rev. D} \textbf{\bibinfo{volume}{102}},
  \bibinfo{pages}{056003} (\bibinfo{year}{2020}), \eprint{2004.06751}.

\bibitem[{\citenamefont{Tanji and Venugopalan}(2017)}]{Tanji:2017suk}
\bibinfo{author}{\bibfnamefont{N.}~\bibnamefont{Tanji}} \bibnamefont{and}
  \bibinfo{author}{\bibfnamefont{R.}~\bibnamefont{Venugopalan}},
  \bibinfo{journal}{Phys. Rev. D} \textbf{\bibinfo{volume}{95}},
  \bibinfo{pages}{094009} (\bibinfo{year}{2017}), \eprint{1703.01372}.

\bibitem[{\citenamefont{Brewer et~al.}(2022)\citenamefont{Brewer,
  Scheihing-Hitschfeld, and Yin}}]{Brewer:2022vkq}
\bibinfo{author}{\bibfnamefont{J.}~\bibnamefont{Brewer}},
  \bibinfo{author}{\bibfnamefont{B.}~\bibnamefont{Scheihing-Hitschfeld}},
  \bibnamefont{and} \bibinfo{author}{\bibfnamefont{Y.}~\bibnamefont{Yin}},
  \bibinfo{journal}{JHEP} \textbf{\bibinfo{volume}{05}}, \bibinfo{pages}{145}
  (\bibinfo{year}{2022}), \eprint{2203.02427}.

\bibitem[{\citenamefont{Ambrus et~al.}(2021)\citenamefont{Ambrus, Busuioc,
  Fotakis, Gallmeister, and Greiner}}]{Ambrus:2021sjg}
\bibinfo{author}{\bibfnamefont{V.~E.} \bibnamefont{Ambrus}},
  \bibinfo{author}{\bibfnamefont{S.}~\bibnamefont{Busuioc}},
  \bibinfo{author}{\bibfnamefont{J.~A.} \bibnamefont{Fotakis}},
  \bibinfo{author}{\bibfnamefont{K.}~\bibnamefont{Gallmeister}},
  \bibnamefont{and} \bibinfo{author}{\bibfnamefont{C.}~\bibnamefont{Greiner}},
  \bibinfo{journal}{Phys. Rev. D} \textbf{\bibinfo{volume}{104}},
  \bibinfo{pages}{094022} (\bibinfo{year}{2021}), \eprint{2102.11785}.

\bibitem[{\citenamefont{Berges et~al.}(2014{\natexlab{a}})\citenamefont{Berges,
  Boguslavski, Schlichting, and Venugopalan}}]{Berges:2013eia}
\bibinfo{author}{\bibfnamefont{J.}~\bibnamefont{Berges}},
  \bibinfo{author}{\bibfnamefont{K.}~\bibnamefont{Boguslavski}},
  \bibinfo{author}{\bibfnamefont{S.}~\bibnamefont{Schlichting}},
  \bibnamefont{and}
  \bibinfo{author}{\bibfnamefont{R.}~\bibnamefont{Venugopalan}},
  \bibinfo{journal}{Phys. Rev. D} \textbf{\bibinfo{volume}{89}},
  \bibinfo{pages}{074011} (\bibinfo{year}{2014}{\natexlab{a}}),
  \eprint{1303.5650}.

\bibitem[{\citenamefont{Berges et~al.}(2014{\natexlab{b}})\citenamefont{Berges,
  Boguslavski, Schlichting, and Venugopalan}}]{Berges:2013fga}
\bibinfo{author}{\bibfnamefont{J.}~\bibnamefont{Berges}},
  \bibinfo{author}{\bibfnamefont{K.}~\bibnamefont{Boguslavski}},
  \bibinfo{author}{\bibfnamefont{S.}~\bibnamefont{Schlichting}},
  \bibnamefont{and}
  \bibinfo{author}{\bibfnamefont{R.}~\bibnamefont{Venugopalan}},
  \bibinfo{journal}{Phys. Rev. D} \textbf{\bibinfo{volume}{89}},
  \bibinfo{pages}{114007} (\bibinfo{year}{2014}{\natexlab{b}}),
  \eprint{1311.3005}.

\bibitem[{\citenamefont{Heller et~al.}(2012)\citenamefont{Heller, Janik, and
  Witaszczyk}}]{Heller:2011ju}
\bibinfo{author}{\bibfnamefont{M.~P.} \bibnamefont{Heller}},
  \bibinfo{author}{\bibfnamefont{R.~A.} \bibnamefont{Janik}}, \bibnamefont{and}
  \bibinfo{author}{\bibfnamefont{P.}~\bibnamefont{Witaszczyk}},
  \bibinfo{journal}{Phys. Rev. Lett.} \textbf{\bibinfo{volume}{108}},
  \bibinfo{pages}{201602} (\bibinfo{year}{2012}), \eprint{1103.3452}.

\bibitem[{\citenamefont{Martinez and Strickland}(2010)}]{Martinez:2010sc}
\bibinfo{author}{\bibfnamefont{M.}~\bibnamefont{Martinez}} \bibnamefont{and}
  \bibinfo{author}{\bibfnamefont{M.}~\bibnamefont{Strickland}},
  \bibinfo{journal}{Nucl. Phys. A} \textbf{\bibinfo{volume}{848}},
  \bibinfo{pages}{183} (\bibinfo{year}{2010}), \eprint{1007.0889}.

\bibitem[{\citenamefont{Florkowski and Ryblewski}(2011)}]{Florkowski:2010cf}
\bibinfo{author}{\bibfnamefont{W.}~\bibnamefont{Florkowski}} \bibnamefont{and}
  \bibinfo{author}{\bibfnamefont{R.}~\bibnamefont{Ryblewski}},
  \bibinfo{journal}{Phys. Rev. C} \textbf{\bibinfo{volume}{83}},
  \bibinfo{pages}{034907} (\bibinfo{year}{2011}), \eprint{1007.0130}.

\bibitem[{\citenamefont{Almaalol et~al.}(2019)\citenamefont{Almaalol,
  Alqahtani, and Strickland}}]{Almaalol:2018jmz}
\bibinfo{author}{\bibfnamefont{D.}~\bibnamefont{Almaalol}},
  \bibinfo{author}{\bibfnamefont{M.}~\bibnamefont{Alqahtani}},
  \bibnamefont{and}
  \bibinfo{author}{\bibfnamefont{M.}~\bibnamefont{Strickland}},
  \bibinfo{journal}{Phys. Rev. C} \textbf{\bibinfo{volume}{99}},
  \bibinfo{pages}{014903} (\bibinfo{year}{2019}), \eprint{1808.07038}.

\bibitem[{\citenamefont{Ferini et~al.}(2009)\citenamefont{Ferini, Colonna,
  Di~Toro, and Greco}}]{Ferini:2008he}
\bibinfo{author}{\bibfnamefont{G.}~\bibnamefont{Ferini}},
  \bibinfo{author}{\bibfnamefont{M.}~\bibnamefont{Colonna}},
  \bibinfo{author}{\bibfnamefont{M.}~\bibnamefont{Di~Toro}}, \bibnamefont{and}
  \bibinfo{author}{\bibfnamefont{V.}~\bibnamefont{Greco}},
  \bibinfo{journal}{Phys. Lett. B} \textbf{\bibinfo{volume}{670}},
  \bibinfo{pages}{325} (\bibinfo{year}{2009}), \eprint{0805.4814}.

\bibitem[{\citenamefont{Plumari et~al.}(2012)\citenamefont{Plumari, Puglisi,
  Scardina, and Greco}}]{Plumari:2012ep}
\bibinfo{author}{\bibfnamefont{S.}~\bibnamefont{Plumari}},
  \bibinfo{author}{\bibfnamefont{A.}~\bibnamefont{Puglisi}},
  \bibinfo{author}{\bibfnamefont{F.}~\bibnamefont{Scardina}}, \bibnamefont{and}
  \bibinfo{author}{\bibfnamefont{V.}~\bibnamefont{Greco}},
  \bibinfo{journal}{Phys. Rev. C} \textbf{\bibinfo{volume}{86}},
  \bibinfo{pages}{054902} (\bibinfo{year}{2012}), \eprint{1208.0481}.

\bibitem[{\citenamefont{Scardina et~al.}(2013)\citenamefont{Scardina, Colonna,
  Plumari, and Greco}}]{Scardina:2012mik}
\bibinfo{author}{\bibfnamefont{F.}~\bibnamefont{Scardina}},
  \bibinfo{author}{\bibfnamefont{M.}~\bibnamefont{Colonna}},
  \bibinfo{author}{\bibfnamefont{S.}~\bibnamefont{Plumari}}, \bibnamefont{and}
  \bibinfo{author}{\bibfnamefont{V.}~\bibnamefont{Greco}},
  \bibinfo{journal}{Phys. Lett. B} \textbf{\bibinfo{volume}{724}},
  \bibinfo{pages}{296} (\bibinfo{year}{2013}), \eprint{1202.2262}.

\bibitem[{\citenamefont{Puglisi et~al.}(2014)\citenamefont{Puglisi, Plumari,
  and Greco}}]{Puglisi:2014sha}
\bibinfo{author}{\bibfnamefont{A.}~\bibnamefont{Puglisi}},
  \bibinfo{author}{\bibfnamefont{S.}~\bibnamefont{Plumari}}, \bibnamefont{and}
  \bibinfo{author}{\bibfnamefont{V.}~\bibnamefont{Greco}},
  \bibinfo{journal}{Phys. Rev. D} \textbf{\bibinfo{volume}{90}},
  \bibinfo{pages}{114009} (\bibinfo{year}{2014}), \eprint{1408.7043}.

\bibitem[{\citenamefont{Scardina et~al.}(2014)\citenamefont{Scardina,
  Perricone, Plumari, Ruggieri, and Greco}}]{Scardina:2014gxa}
\bibinfo{author}{\bibfnamefont{F.}~\bibnamefont{Scardina}},
  \bibinfo{author}{\bibfnamefont{D.}~\bibnamefont{Perricone}},
  \bibinfo{author}{\bibfnamefont{S.}~\bibnamefont{Plumari}},
  \bibinfo{author}{\bibfnamefont{M.}~\bibnamefont{Ruggieri}}, \bibnamefont{and}
  \bibinfo{author}{\bibfnamefont{V.}~\bibnamefont{Greco}},
  \bibinfo{journal}{Phys. Rev. C} \textbf{\bibinfo{volume}{90}},
  \bibinfo{pages}{054904} (\bibinfo{year}{2014}), \eprint{1408.1313}.

\bibitem[{\citenamefont{Plumari
  et~al.}(2015{\natexlab{b}})\citenamefont{Plumari, Guardo, Scardina, and
  Greco}}]{Plumari:2015cfa}
\bibinfo{author}{\bibfnamefont{S.}~\bibnamefont{Plumari}},
  \bibinfo{author}{\bibfnamefont{G.~L.} \bibnamefont{Guardo}},
  \bibinfo{author}{\bibfnamefont{F.}~\bibnamefont{Scardina}}, \bibnamefont{and}
  \bibinfo{author}{\bibfnamefont{V.}~\bibnamefont{Greco}},
  \bibinfo{journal}{Phys. Rev. C} \textbf{\bibinfo{volume}{92}},
  \bibinfo{pages}{054902} (\bibinfo{year}{2015}{\natexlab{b}}),
  \eprint{1507.05540}.

\bibitem[{\citenamefont{Scardina et~al.}(2017)\citenamefont{Scardina, Das,
  Minissale, Plumari, and Greco}}]{Scardina:2017ipo}
\bibinfo{author}{\bibfnamefont{F.}~\bibnamefont{Scardina}},
  \bibinfo{author}{\bibfnamefont{S.~K.} \bibnamefont{Das}},
  \bibinfo{author}{\bibfnamefont{V.}~\bibnamefont{Minissale}},
  \bibinfo{author}{\bibfnamefont{S.}~\bibnamefont{Plumari}}, \bibnamefont{and}
  \bibinfo{author}{\bibfnamefont{V.}~\bibnamefont{Greco}},
  \bibinfo{journal}{Phys. Rev. C} \textbf{\bibinfo{volume}{96}},
  \bibinfo{pages}{044905} (\bibinfo{year}{2017}), \eprint{1707.05452}.

\bibitem[{\citenamefont{Sun et~al.}(2020)\citenamefont{Sun, Plumari, and
  Greco}}]{Sun:2019gxg}
\bibinfo{author}{\bibfnamefont{Y.}~\bibnamefont{Sun}},
  \bibinfo{author}{\bibfnamefont{S.}~\bibnamefont{Plumari}}, \bibnamefont{and}
  \bibinfo{author}{\bibfnamefont{V.}~\bibnamefont{Greco}},
  \bibinfo{journal}{Eur. Phys. J. C} \textbf{\bibinfo{volume}{80}},
  \bibinfo{pages}{16} (\bibinfo{year}{2020}), \eprint{1907.11287}.

\bibitem[{\citenamefont{Sambataro et~al.}(2020)\citenamefont{Sambataro,
  Plumari, and Greco}}]{Sambataro:2020pge}
\bibinfo{author}{\bibfnamefont{M.~L.} \bibnamefont{Sambataro}},
  \bibinfo{author}{\bibfnamefont{S.}~\bibnamefont{Plumari}}, \bibnamefont{and}
  \bibinfo{author}{\bibfnamefont{V.}~\bibnamefont{Greco}},
  \bibinfo{journal}{Eur. Phys. J. C} \textbf{\bibinfo{volume}{80}},
  \bibinfo{pages}{1140} (\bibinfo{year}{2020}), \eprint{2005.14470}.

\bibitem[{\citenamefont{Gabbana et~al.}(2020)\citenamefont{Gabbana, Plumari,
  Galesi, Greco, Simeoni, Succi, and Tripiccione}}]{Gabbana:2019uqv}
\bibinfo{author}{\bibfnamefont{A.}~\bibnamefont{Gabbana}},
  \bibinfo{author}{\bibfnamefont{S.}~\bibnamefont{Plumari}},
  \bibinfo{author}{\bibfnamefont{G.}~\bibnamefont{Galesi}},
  \bibinfo{author}{\bibfnamefont{V.}~\bibnamefont{Greco}},
  \bibinfo{author}{\bibfnamefont{D.}~\bibnamefont{Simeoni}},
  \bibinfo{author}{\bibfnamefont{S.}~\bibnamefont{Succi}}, \bibnamefont{and}
  \bibinfo{author}{\bibfnamefont{R.}~\bibnamefont{Tripiccione}},
  \bibinfo{journal}{Phys. Rev. C} \textbf{\bibinfo{volume}{101}},
  \bibinfo{pages}{064904} (\bibinfo{year}{2020}), \eprint{1912.10455}.

\bibitem[{\citenamefont{Wong}(1982)}]{Wong:1982zzb}
\bibinfo{author}{\bibfnamefont{C.-Y.} \bibnamefont{Wong}},
  \bibinfo{journal}{Phys. Rev. C} \textbf{\bibinfo{volume}{25}},
  \bibinfo{pages}{1460} (\bibinfo{year}{1982}).

\bibitem[{\citenamefont{Lang et~al.}(1993)\citenamefont{Lang, Babovsky,
  Cassing, Mosel, Reusch, and Weber}}]{LANG1993391}
\bibinfo{author}{\bibfnamefont{A.}~\bibnamefont{Lang}},
  \bibinfo{author}{\bibfnamefont{H.}~\bibnamefont{Babovsky}},
  \bibinfo{author}{\bibfnamefont{W.}~\bibnamefont{Cassing}},
  \bibinfo{author}{\bibfnamefont{U.}~\bibnamefont{Mosel}},
  \bibinfo{author}{\bibfnamefont{H.-G.} \bibnamefont{Reusch}},
  \bibnamefont{and} \bibinfo{author}{\bibfnamefont{K.}~\bibnamefont{Weber}},
  \bibinfo{journal}{Journal of Computational Physics}
  \textbf{\bibinfo{volume}{106}}, \bibinfo{pages}{391} (\bibinfo{year}{1993}),
  ISSN \bibinfo{issn}{0021-9991},
  \urlprefix\url{https://www.sciencedirect.com/science/article/pii/S0021999183711162}.

\bibitem[{\citenamefont{Kovtun et~al.}(2005)\citenamefont{Kovtun, Son, and
  Starinets}}]{Kovtun:2004de}
\bibinfo{author}{\bibfnamefont{P.}~\bibnamefont{Kovtun}},
  \bibinfo{author}{\bibfnamefont{D.~T.} \bibnamefont{Son}}, \bibnamefont{and}
  \bibinfo{author}{\bibfnamefont{A.~O.} \bibnamefont{Starinets}},
  \bibinfo{journal}{Phys. Rev. Lett.} \textbf{\bibinfo{volume}{94}},
  \bibinfo{pages}{111601} (\bibinfo{year}{2005}), \eprint{hep-th/0405231}.

\bibitem[{\citenamefont{{Denicol} and {Rischke}}(2021)}]{Denicol:book}
\bibinfo{author}{\bibfnamefont{G.~S.} \bibnamefont{{Denicol}}}
  \bibnamefont{and} \bibinfo{author}{\bibfnamefont{D.~H.}
  \bibnamefont{{Rischke}}}, \emph{\bibinfo{title}{{Microscopic Foundations of
  Relativistic Fluid Dynamics}}} (\bibinfo{year}{2021}).

\bibitem[{\citenamefont{Strickland and Tantary}(2019)}]{Strickland:2019hff}
\bibinfo{author}{\bibfnamefont{M.}~\bibnamefont{Strickland}} \bibnamefont{and}
  \bibinfo{author}{\bibfnamefont{U.}~\bibnamefont{Tantary}},
  \bibinfo{journal}{JHEP} \textbf{\bibinfo{volume}{10}}, \bibinfo{pages}{069}
  (\bibinfo{year}{2019}), \eprint{1903.03145}.

\bibitem[{\citenamefont{Denicol et~al.}(2010)\citenamefont{Denicol, Koide, and
  Rischke}}]{Denicol:2010xn}
\bibinfo{author}{\bibfnamefont{G.~S.} \bibnamefont{Denicol}},
  \bibinfo{author}{\bibfnamefont{T.}~\bibnamefont{Koide}}, \bibnamefont{and}
  \bibinfo{author}{\bibfnamefont{D.~H.} \bibnamefont{Rischke}},
  \bibinfo{journal}{Phys. Rev. Lett.} \textbf{\bibinfo{volume}{105}},
  \bibinfo{pages}{162501} (\bibinfo{year}{2010}), \eprint{1004.5013}.

\bibitem[{\citenamefont{Romatschke and Strickland}(2003)}]{Romatschke:2003ms}
\bibinfo{author}{\bibfnamefont{P.}~\bibnamefont{Romatschke}} \bibnamefont{and}
  \bibinfo{author}{\bibfnamefont{M.}~\bibnamefont{Strickland}},
  \bibinfo{journal}{Phys. Rev. D} \textbf{\bibinfo{volume}{68}},
  \bibinfo{pages}{036004} (\bibinfo{year}{2003}), \eprint{hep-ph/0304092}.

\bibitem[{\citenamefont{Romatschke}(2012)}]{Romatschke:2011qp}
\bibinfo{author}{\bibfnamefont{P.}~\bibnamefont{Romatschke}},
  \bibinfo{journal}{Phys. Rev. D} \textbf{\bibinfo{volume}{85}},
  \bibinfo{pages}{065012} (\bibinfo{year}{2012}), \eprint{1108.5561}.

\bibitem[{\citenamefont{Alqahtani et~al.}(2018)\citenamefont{Alqahtani,
  Nopoush, and Strickland}}]{Alqahtani:2017mhy}
\bibinfo{author}{\bibfnamefont{M.}~\bibnamefont{Alqahtani}},
  \bibinfo{author}{\bibfnamefont{M.}~\bibnamefont{Nopoush}}, \bibnamefont{and}
  \bibinfo{author}{\bibfnamefont{M.}~\bibnamefont{Strickland}},
  \bibinfo{journal}{Prog. Part. Nucl. Phys.} \textbf{\bibinfo{volume}{101}},
  \bibinfo{pages}{204} (\bibinfo{year}{2018}), \eprint{1712.03282}.

\bibitem[{\citenamefont{Bjorken}(1983)}]{Bjorken:1982qr}
\bibinfo{author}{\bibfnamefont{J.~D.} \bibnamefont{Bjorken}},
  \bibinfo{journal}{Phys. Rev. D} \textbf{\bibinfo{volume}{27}},
  \bibinfo{pages}{140} (\bibinfo{year}{1983}).

\bibitem[{\citenamefont{Wagner et~al.}(2023)\citenamefont{Wagner, Ambrus, and
  Molnar}}]{Wagner:2023joq}
\bibinfo{author}{\bibfnamefont{D.}~\bibnamefont{Wagner}},
  \bibinfo{author}{\bibfnamefont{V.~E.} \bibnamefont{Ambrus}},
  \bibnamefont{and} \bibinfo{author}{\bibfnamefont{E.}~\bibnamefont{Molnar}}
  (\bibinfo{year}{2023}), \eprint{2309.09335}.

\bibitem[{\citenamefont{Jankowski and Spali\'nski}(2023)}]{Jankowski:2023fdz}
\bibinfo{author}{\bibfnamefont{J.}~\bibnamefont{Jankowski}} \bibnamefont{and}
  \bibinfo{author}{\bibfnamefont{M.}~\bibnamefont{Spali\'nski}},
  \bibinfo{journal}{Prog. Part. Nucl. Phys.} \textbf{\bibinfo{volume}{132}},
  \bibinfo{pages}{104048} (\bibinfo{year}{2023}), \eprint{2303.09414}.

\bibitem[{\citenamefont{Giacalone et~al.}(2019)\citenamefont{Giacalone,
  Mazeliauskas, and Schlichting}}]{Giacalone:2019ldn}
\bibinfo{author}{\bibfnamefont{G.}~\bibnamefont{Giacalone}},
  \bibinfo{author}{\bibfnamefont{A.}~\bibnamefont{Mazeliauskas}},
  \bibnamefont{and}
  \bibinfo{author}{\bibfnamefont{S.}~\bibnamefont{Schlichting}},
  \bibinfo{journal}{Phys. Rev. Lett.} \textbf{\bibinfo{volume}{123}},
  \bibinfo{pages}{262301} (\bibinfo{year}{2019}), \eprint{1908.02866}.

\bibitem[{\citenamefont{Huovinen and Molnar}(2009)}]{Huovinen:2008te}
\bibinfo{author}{\bibfnamefont{P.}~\bibnamefont{Huovinen}} \bibnamefont{and}
  \bibinfo{author}{\bibfnamefont{D.}~\bibnamefont{Molnar}},
  \bibinfo{journal}{Phys. Rev. C} \textbf{\bibinfo{volume}{79}},
  \bibinfo{pages}{014906} (\bibinfo{year}{2009}), \eprint{0808.0953}.

\bibitem[{\citenamefont{Betz et~al.}(2011)\citenamefont{Betz, Denicol, Koide,
  Molnar, Niemi, and Rischke}}]{Betz:2010cx}
\bibinfo{author}{\bibfnamefont{B.}~\bibnamefont{Betz}},
  \bibinfo{author}{\bibfnamefont{G.~S.} \bibnamefont{Denicol}},
  \bibinfo{author}{\bibfnamefont{T.}~\bibnamefont{Koide}},
  \bibinfo{author}{\bibfnamefont{E.}~\bibnamefont{Molnar}},
  \bibinfo{author}{\bibfnamefont{H.}~\bibnamefont{Niemi}}, \bibnamefont{and}
  \bibinfo{author}{\bibfnamefont{D.~H.} \bibnamefont{Rischke}},
  \bibinfo{journal}{EPJ Web Conf.} \textbf{\bibinfo{volume}{13}},
  \bibinfo{pages}{07005} (\bibinfo{year}{2011}), \eprint{1012.5772}.

\bibitem[{\citenamefont{Denicol et~al.}(2011)\citenamefont{Denicol, Noronha,
  Niemi, and Rischke}}]{Denicol:2011fa}
\bibinfo{author}{\bibfnamefont{G.~S.} \bibnamefont{Denicol}},
  \bibinfo{author}{\bibfnamefont{J.}~\bibnamefont{Noronha}},
  \bibinfo{author}{\bibfnamefont{H.}~\bibnamefont{Niemi}}, \bibnamefont{and}
  \bibinfo{author}{\bibfnamefont{D.~H.} \bibnamefont{Rischke}},
  \bibinfo{journal}{Phys. Rev. D} \textbf{\bibinfo{volume}{83}},
  \bibinfo{pages}{074019} (\bibinfo{year}{2011}), \eprint{1102.4780}.

\bibitem[{\citenamefont{Romatschke}(2018)}]{Romatschke:2017vte}
\bibinfo{author}{\bibfnamefont{P.}~\bibnamefont{Romatschke}},
  \bibinfo{journal}{Phys. Rev. Lett.} \textbf{\bibinfo{volume}{120}},
  \bibinfo{pages}{012301} (\bibinfo{year}{2018}), \eprint{1704.08699}.

\bibitem[{\citenamefont{Alalawi and Strickland}(2022)}]{Alalawi:2022pmg}
\bibinfo{author}{\bibfnamefont{H.}~\bibnamefont{Alalawi}} \bibnamefont{and}
  \bibinfo{author}{\bibfnamefont{M.}~\bibnamefont{Strickland}},
  \bibinfo{journal}{JHEP (2022)} \textbf{\bibinfo{volume}{12}},
  \bibinfo{pages}{143} (\bibinfo{year}{2022}), \eprint{2210.00658}.

\bibitem[{\citenamefont{Csernai et~al.}(2006)\citenamefont{Csernai, Kapusta,
  and McLerran}}]{Csernai:2006zz}
\bibinfo{author}{\bibfnamefont{L.~P.} \bibnamefont{Csernai}},
  \bibinfo{author}{\bibfnamefont{J.~I.} \bibnamefont{Kapusta}},
  \bibnamefont{and} \bibinfo{author}{\bibfnamefont{L.~D.}
  \bibnamefont{McLerran}}, \bibinfo{journal}{Phys. Rev. Lett.}
  \textbf{\bibinfo{volume}{97}}, \bibinfo{pages}{152303}
  (\bibinfo{year}{2006}), \eprint{nucl-th/0604032}.

\bibitem[{\citenamefont{Plumari et~al.}(2013)\citenamefont{Plumari, Greco, and
  Csernai}}]{Plumari:2013bga}
\bibinfo{author}{\bibfnamefont{S.}~\bibnamefont{Plumari}},
  \bibinfo{author}{\bibfnamefont{V.}~\bibnamefont{Greco}}, \bibnamefont{and}
  \bibinfo{author}{\bibfnamefont{L.~P.} \bibnamefont{Csernai}}
  (\bibinfo{year}{2013}), \eprint{1304.6566}.

\bibitem[{\citenamefont{Yang and Chen}(2023)}]{Yang:2022ixy}
\bibinfo{author}{\bibfnamefont{Z.}~\bibnamefont{Yang}} \bibnamefont{and}
  \bibinfo{author}{\bibfnamefont{L.-W.} \bibnamefont{Chen}},
  \bibinfo{journal}{Phys. Rev. C} \textbf{\bibinfo{volume}{107}},
  \bibinfo{pages}{064910} (\bibinfo{year}{2023}), \eprint{2207.13534}.

\bibitem[{\citenamefont{Meyer}(2007)}]{Meyer:2007ic}
\bibinfo{author}{\bibfnamefont{H.~B.} \bibnamefont{Meyer}},
  \bibinfo{journal}{Phys. Rev. D} \textbf{\bibinfo{volume}{76}},
  \bibinfo{pages}{101701} (\bibinfo{year}{2007}), \eprint{0704.1801}.

\bibitem[{\citenamefont{{Bors\'anyi, Sz. and Fodor, Zoltan and Giordano, Matteo
  and Katz, Sandor D. and Pasztor, Attila and Ratti, Claudia and Sch\"afer,
  Andreas and Szabo, Kalman K. and T\'oth, Balint\,
  C.}}(2018)}]{Borsanyi:2018srz}
\bibinfo{author}{\bibnamefont{{Bors\'anyi, Sz. and Fodor, Zoltan and Giordano,
  Matteo and Katz, Sandor D. and Pasztor, Attila and Ratti, Claudia and
  Sch\"afer, Andreas and Szabo, Kalman K. and T\'oth, Balint\, C.}}},
  \bibinfo{journal}{Phys. Rev. D} \textbf{\bibinfo{volume}{98}},
  \bibinfo{pages}{014512} (\bibinfo{year}{2018}), \eprint{1802.07718}.

\bibitem[{\citenamefont{Chen et~al.}(2007)\citenamefont{Chen, Li, Liu, and
  Nakano}}]{Chen:2007xe}
\bibinfo{author}{\bibfnamefont{J.-W.} \bibnamefont{Chen}},
  \bibinfo{author}{\bibfnamefont{Y.-H.} \bibnamefont{Li}},
  \bibinfo{author}{\bibfnamefont{Y.-F.} \bibnamefont{Liu}}, \bibnamefont{and}
  \bibinfo{author}{\bibfnamefont{E.}~\bibnamefont{Nakano}},
  \bibinfo{journal}{Phys. Rev. D} \textbf{\bibinfo{volume}{76}},
  \bibinfo{pages}{114011} (\bibinfo{year}{2007}), \eprint{hep-ph/0703230}.

\bibitem[{\citenamefont{Bernhard et~al.}(2019)\citenamefont{Bernhard, Moreland,
  and Bass}}]{Bernhard:2019bmu}
\bibinfo{author}{\bibfnamefont{J.~E.} \bibnamefont{Bernhard}},
  \bibinfo{author}{\bibfnamefont{J.~S.} \bibnamefont{Moreland}},
  \bibnamefont{and} \bibinfo{author}{\bibfnamefont{S.~A.} \bibnamefont{Bass}},
  \bibinfo{journal}{Nature Phys.} \textbf{\bibinfo{volume}{15}},
  \bibinfo{pages}{1113} (\bibinfo{year}{2019}).

\end{thebibliography}

\end{document}